\documentclass[a4paper,11pt,reqno]{amsart}

\usepackage{amssymb, amsmath, bm}
\usepackage{mathrsfs}
\usepackage{dsfont}
\usepackage{appendix}

\allowdisplaybreaks

\newtheorem{thm}{Theorem}[section]
\newtheorem{prp}[thm]{Proposition}
\newtheorem{lem}[thm]{Lemma}
\newtheorem{dfn}[thm]{Definition}

\newtheorem{cor}[thm]{Corollary}
\newtheorem{example}[thm]{Example}
\newenvironment{exa}{\begin{example} \rm }{ \end{example}}
\newtheorem{remark}[thm]{Remark}

\newenvironment{rmk}{\begin{remark} \rm }{\hfill $\Box$ \end{remark}}
\newenvironment{prf}{\noindent {\it Proof.}  }{\hfill $\Box$}
\newenvironment{prfof}[1]{\noindent {\it Proof of #1.} \ }{\hfill $\Box$}

\newcommand{\al}{\alpha}
\newcommand{\Ta}{\Theta} \newcommand{\ta}{\theta}

\newcommand{\ga}{\gamma}
\newcommand{\Ga}{\Gamma}
\newcommand{\ep}{\epsilon}
\newcommand{\ka}{\kappa}

\newcommand{\la}{\lambda}

\newcommand{\Om}{\Omega}

\newcommand{\de}{\delta}

\newcommand{\pa}{\partial}
\newcommand{\tr}{{\rm tr\/}}

\newcommand{\ve}{\varepsilon}

\newcommand{\fh}{\mathfrak{h}}
\newcommand\fg{\mathfrak{g}}
\newcommand\Z{\mathbb{Z}}
\newcommand\C{\mathbb{C}}

\newcommand\vac{|0\rangle}
\newcommand\fsl{\mathfrak{sl}}
\newcommand{\id}{\mathrm{Id}}
\newcommand{\res}{\mathrm{Res}}

\newcommand{\bs}[1]{\boldsymbol{#1}}

\newcommand{\btimes}{\boldsymbol{\otimes}}
\newcommand{\tor}{\mathrm{tor}}

\begin{document}

\title[]{Integrable hierarchy for homogeneous realization of the toroidal Lie algebra $\mathcal{L}^{\tor}_{r+1}(\fsl_\ell)$}
\author[]{Chao-Zhong Wu, Yi Yang$^*$}
\thanks{*Corresponding author.}
\dedicatory {School of Mathematics, Sun Yat-sen University\\
Guangzhou, 510000, China  \\
Email address:  wuchaozhong@sysu.edu.cn, yangy875@mail2.sysu.edu.cn}
\begin{abstract}
Starting from a fairly explicit homogeneous realization of the toroidal Lie algebra $\mathcal{L}^{\tor}_{r+1}(\fsl_\ell)$ via a lattice vertex algebra, we derive an integrable hierarchy of Hirota bilinear equations. Moreover, we represent this hierarchy in the form of Lax equations, and show that it is an extension of a certain reduction of the $\ell$-component KP hierarchy.
\end{abstract}
\maketitle

\section{Introduction}

An important topic in the theory of integrable systems is its relationship with affine Lie algebras and their representations, see \cite{DS,Jimbo1983,Kac1990,KvdL-nKP,KRR,KW} for example. Among the fruitful research results on this topic, many integrable systems are constructed or represented in the form of Hirota bilinear equations. As a generalization of untwisted affine Lie algebras, a toroidal Lie algebra is the central extension of the tensor product of a simple Lie algebra (only complex Lie algebras are considered below) and a certain  ring of Laurent polynomials in multi-variables \cite{MRY}. A natural question is how to employ toroidal Lie algebras as well as their representations (see for instance \cite{Bakalov2004,Bakalov2021,BBS,Billig1998,BL,ERM,FM,Tan}) to investigate integrable hierarchies.

In late 1990s, Billig \cite{Billig1999} showed how to construct hierarchies of PDEs for the principal vertex operator realization of a general toroidal Lie algebra with multi-variables, while Iohara, Saito and Wakimoto \cite{ISW} derived Hirota bilinear equations from the homogeneous and the principal vertex operator realizations of $2$-toroidal Lie algebras of ADE type. Such constructions rely essentially on certain generalized Casimir operators (cf.\cite{KW}) with technical computations, and give explicit equations in the case the $2$-toroidal Lie algebra of $\fsl_2$ ($\mathcal{L}^{\tor}_2(\fsl_2)$ for short). In particular, the hierarchy for $\mathcal{L}^{\tor}_2(\fsl_2)$ obtained in \cite{Billig1999} is an extension of the Korteweg-de Vries (KdV) hierarchy, which is well-known associated to the simple Lie algebra $\fsl_2$ or its affinization, say, $\mathcal{L}(\fsl_2)$ (also written as $\mathcal{L}^{\tor}_1(\fsl_2)$ in Appendix~A below).  Note that Lax representations of these Hirota bilinear equations have not been considered in \cite{Billig1999,ISW}.

In early 2000s, Ikeda, Kakei and Takasaki \cite{IT,KIT} constructed integrable hierarchies associated to toroidal Lie algebras in an approach of boson-fermion correspondence (cf.\cite{Jimbo1983}). Along this line, the integrable hierarchy derived from the homogeneous realization of  $\mathcal{L}^{\tor}_2(\fsl_2)$, named as the hierarchy of the (2+1)-dimensional nonlinear Schr\"odinger equation in \cite{IT}, was shown to be a reduction of the two-component KP hierarchy and related to the self-dual Yang-Mills equation. Moreover, the integrable hierarchy derived in \cite{IT} from the principal realization of  $\mathcal{L}^{\tor}_2(\fsl_\ell)$ was shown to be the $\ell$-Bogoyavlensky hierarchy (cf.\cite{Bog,IT}).

In 2010, Tan \cite{TanY} deduced integrable systems of Toda type associated to  toroidal Lie algebras for simple Lie algebras via zero-curvature equations, and he obtained a Hamiltonian formalism for them in truncated cases.

In the representation theory of toroidal Lie algebras, the notion of lattice vertex algebra \cite{Bakalov2004,Kac1997} plays an important role; see \cite{Bakalov2021,BBS,ERM} and references therein. From this point of view, in the present note we want to study integrable hierarchies associated to toroidal Lie algebras in a way distinct to \cite{Billig1999,ISW,IT,KIT}. More exactly, we will write down explicitly the homogeneous (untwisted) realization obtained in \cite{Bakalov2021,BBS} of the toroidal Lie algebra $\mathcal{L}^{\tor}_{r+1}(\fsl_\ell)$ with an arbitrary positive integer $r$ via a certain lattice vertex algebra, then derive a hierarchy of Hirota bilinear equations satisfied by the vacuum orbit of the Lie group of $\mathcal{L}^{\tor}_{r+1}(\fsl_\ell)$. Moreover, we will represent these Hirota bilinear equations in the form of Lax equations of matrix-coefficient pseudo-differential operators, and show that this hierarchy is an extension of the $(1,1,\dots,1)$-reduction of the $\ell$-component KP hierarchy \cite{KvdL-nKP,KvdL-KPred}. Our main results will be given in Theorems~\ref{thm-bltor} and \ref{thm-Laxtor} below. We remark that, whenever $r=1$ and $\ell=2$, our results agree with that for the hierarchy given in \cite{KIT} for $\mathcal{L}^{\tor}_{2}(\fsl_2)$.

This note is arranged as follows. In the next section, we will recall the concept of lattice vertex algebra. In Section~3, we will see that a subalgebra of a certain lattice vertex algebra is isomorphic to the affine Lie algebra $\mathcal{L}(\fsl_\ell)$ of type $A_{\ell-1}^{(1)}$, and that the corresponding integrable hierarchy is a reduction of the $\ell$-component KP hierarchy. In Section~4, starting from an extension of the above lattice vertex algebra, we write down a homogeneous realization of the toroidal Lie algebra $\mathcal{L}^{\tor}_{r+1}(\fsl_\ell)$, then prove our main results. Section~5 is dedicated to some remarks. Finally, in order to make this note self-contained in a sense, we will recall the definition of a toroidal Lie algebra in Appendix~A, and verify Proposition~\ref{thm-hep} straightforwardly in Appendix~B.

\section{Lattice vertex algebras}\label{latticealge}

Let us review the definition and some properties of lattice vertex (super)algebras, following mainly the notions of \cite{Kac1997}. In what follows the notions lattice vertex superalgebra and lattice vertex algebra will not be distinguished, such as in \cite{Kac1997}.

We assume $Q$ to be a free abelian group of rank $\ell$, which is endowed with a $\mathbb{Z}$-valued nondegenerate symmetric bilinear form $(\cdot|\cdot)$.
The bilinear form is naturally extended to the complexification of $Q$, say,
$\mathfrak{h}=\mathbb{C}\otimes_{\mathbb{Z}}Q$.
Consider $\mathfrak{h}$ as a commutative Lie algebra, and its affinization reads
\[
\widehat{\mathfrak{h}}=\mathfrak{h}[s, s^{-1}]\oplus\mathbb{C}K.
\]
Here $s$ is a parameter, $K$ is the central element, and the Lie bracket is defined by
  \begin{align}
  [h s^m +\la K, h' s^n+\mu K]=m\delta_{m+n,0}(h|h')K, \label{Heisenberg}
  \end{align}
for any $h,h'\in \fh$; $m,n\in\mathbb{Z}$ and $\la, \mu\in\mathbb{C}$.

Let
\[
\varepsilon:Q\times Q\longrightarrow\{\pm1\}
\]
be a  bi-multiplicative function. Namely, for any $\al, \beta, \ga\in Q$, one has
\begin{equation}\label{bimulti}
\ve(\al+\beta,\ga)=\ve(\al,\ga)\ve(\beta,\ga), \quad \ve(\al, \beta+\ga)=\ve(\al,\beta)\ve(\al,\ga).
\end{equation}
By using this bi-multiplicative function, one introduces a twisted group algebra $\mathbb{C}_{\varepsilon}[Q]$ with a basis $\{e^{\alpha}\}_{\alpha\in Q}$ and multiplication defined by
\[
e^{\alpha}e^{\beta} =\varepsilon(\alpha,\beta)e^{\alpha+\beta}, \quad \al, \beta\in Q.
\]
Clearly, this is an associative algebra.

Let $S$ be the symmetric algebra over the space $\fh^{<0}=\sum_{j<0}\fh s^{j}$.
The lattice vertex algebra associated to $Q$ is a vertex algebra such that
\begin{itemize}
\item the space of states is
\begin{equation}
V_Q=S\otimes\mathbb{C}_{\varepsilon}[Q]
\end{equation}
with the parity
\begin{equation}\label{parity}
{\mathrm p}\left(g\otimes e^{\beta}\right)=(\beta|\beta)\in \Z/2\Z \quad \hbox{for any} \quad g\in S, ~\beta\in Q;
\end{equation}
\item the vacuum vector is $|0\rangle=1\otimes e^0$;
\item the state-field correspondence is a linear map
\begin{equation}\label{}
\mathrm{Y}(\cdot,z) : V_Q \rightarrow {\rm End}(V_Q)[[z,z^{-1}]]
\end{equation}
defined as follows.
\end{itemize}
Firstly, one has
\[
\mathrm{Y}\left(|0\rangle,z\right)=\mathrm{Id}_{V_Q}.
\]
Given any $h\in\fh$ and $m\in\Z$, one introduces a linear transformation $h_{(-m)}\in {\rm End}(V_Q)$ by
\begin{align}\label{hmm}
h_{(-m)}\left(g\otimes e^{\beta}\right)=\left\{
    \begin{array}{cl}
            h s^{-m}g\otimes e^{\beta}, &  m>0;\\
            (h|\beta)g\otimes e^{\beta}, &  m=0; \\
      \sum\limits_{i=1}^k\delta_{-m,n_i}n_i(h|g_i)g_1 s^{-n_1}\dots \widehat{g_{i} s^{-n_{i}}}  \dots   g_k s^{-n_k}\otimes e^{\beta}, &  m<0, \\
    \end{array}
  \right.
\end{align}
where $\beta\in Q$, $g=g_1 s^{-n_1}\cdot \dots \cdot g_k s^{-n_k}\in S$ with $g_i\in \fh$ and $n_i\in\Z_{>0}$, and $\widehat{g_{i} s^{-n_{i}}}$ means that the $i$-th factor is dropped. In particular, one has
\[
h_{(-m)}\left(1\otimes e^{\beta}\right)=0, \quad m\in\Z_{<0}.
\]
For $h\in\fh$ and $\al\in Q\subset\fh$, denote
\begin{align}\label{hz}
h(z)=&\sum_{m\in\Z} h_{(-m)}z^{m-1}, \\
E^\al(z)=& e^{\alpha}z^{\alpha_{(0)}}\exp\left(\sum_{n>0}\alpha_{(-n)}\frac{z^{n}}{n}\right){\rm exp}\left(\sum_{n<0}\alpha_{(-n)}\frac{z^{n}}{n}\right),\label{ealz}
\end{align}
where the action of $e^{\al}$ on $V_Q$ is given by
\begin{align*}
 e^{\alpha}\left(g\otimes e^{\beta}\right)=\varepsilon(\alpha,\beta)g\otimes e^{\alpha+\beta}.
\end{align*}
Then for $\al\in Q$, $h_1,\dots,h_k\in\fh$ and $m_1,\dots,m_k\in\Z_{\ge0}$, one assigns
\begin{align}
&\mathrm{Y}\left( h_{1}s^{-m_1-1}\cdot \dots \cdot h_{k}s^{-m_k-1}\otimes e^{\alpha},z\right) \nonumber\\
=&\frac{1}{m_1!\dots m_n!} :\partial_z^{m_1}h_1(z)\cdot \dots\cdot\partial_z^{m_k}h_k(z) \cdot E^\al(z):. \label{Yhhea}
\end{align}
Here the normal order product ``$: ~:$'' is expanded from the right to the left such that, for any formal series $A(z)=\sum_{n\in\Z}A_n z^n$ one has
\[
:\pa_z^{m_i} h_i(z)\cdot A(z):=\left( \pa_z^{m_i} h_i(z)\right)_{\ge0} A(z) + A(z) \left( \pa_z^{m_i} h_i(z)\right)_{<0},
\]
with $A(z)_{\ge0}=\sum_{n\ge0}A_{n}z^{n}$ and $A(z)_{<0}=\sum_{n<0}A_{n}z^{n}$.

For any $a\in V_Q$, we write
\begin{equation}\label{}
\mathrm{Y}(a,z)=\sum_{m\in\mathbb{Z}}a_{(m)}z^{-m-1}.
\end{equation}
For instance, given $h\in\fh$ and $\al\in Q$, one has 
\begin{align*}
\mathrm{Y}\left(h_{(-1)}\vac,z\right)=h(z), \quad \mathrm{Y}\left(e^{\al},z\right)=E^\al(z).
\end{align*}
For $a\in V_Q$ one sees that
\begin{equation}\label{amvac}
a_{(m)}\vac=0, \quad m\ge0,
\end{equation}
and that
\[
a_{(-1)}\vac=a.
\]
From now on, when there is no confusion we will use $h\in \mathfrak{h}$ and $e^\al$ ($\al\in Q$) to stand for the elements $h_{(-1)}\vac$ and $1\otimes e^\al$, respectively.

Given  $a,b\in V_Q$, it is known that
\begin{align}
 [a_{(m)},b_{(n)}]=\sum_{j\geq0}\binom{m}{j}(a_{(j)}b)_{(m+n-j)}, \quad m,n\in \Z, \label{ambnbra}
 \end{align}
where
\[
\binom{m}{j}=\frac{m(m-1)\dots(m-j+1)}{j!}.
\]
In other words,
\begin{align}
 [\mathrm{Y}(a,z),\mathrm{Y}(b,w)]=\sum_{j\geq0}\mathrm{Y}(a_{(j)}b,w)\frac{\pa^j_w\delta(z-w)}{j!},\label{bracket}
 \end{align}
where $\delta(z,w)=z^{-1}\sum_{n\in\mathbb{Z}}\left(w/z\right)^n$.

\section{From the affine Lie algebra $\mathcal{L}(\fsl_{\ell})$ to Hirota bilinear equations}\label{slireal}

It is known that an affine Lie algebra of ADE type can be realized faithfully as a subalgebra of a certain lattice vertex algebra associated to the corresponding root lattice, see for example \cite{FLM,Kac1997}. Instead of the elegant settings that appear in {\it loc. cit.}, in this section we want write down a lattice vertex algebra realization of the affine Lie algebra $\mathcal{L}(\fsl_{\ell})$ in a fairly explicit way. Based on such an explicit realization, it can be derived a system of Hirota bilinear equations that is equivalent to a certain reduction of the multi-component KP hierarchy. We do not claim any credit for the results presented in this section, which can be found in for example \cite{FLM,Kac1997,KRR,KvdL-KPred}. Just for the convenience of the readers and preparation for the next section, brief proofs of partial results will be given in a straightforward way.

\subsection{Lattice vertex algebras and the affine Lie algebra $\mathcal{L}(\fsl_{\ell})$}
From now on we assume $Q=\oplus_{i=1}^\ell \Z v_i$ to be the rank-$\ell$ lattice, whose bilinear form is given by
\[
(v_i|v_j)=\delta_{i,j}, \quad 1\le i,j\le\ell.
\]
Moreover, on $Q$ it is fixed a bi-multiplication function by $\ve(v_i,v_j)=\ep_{i j}$ with
\begin{equation}\label{}
\ep_{i j}=\left\{
\begin{array}{cl}
               1, & i<j; \\
               -1, & i\ge j.
             \end{array}
             \right.
\end{equation}
Here and below, the subscript of $v_i$ runs from $1$ to $\ell$ unless otherwise stated.
Then as above in the previous section we have the lattice vertex algebra $V_Q$ associated to $Q$.

Consider the following elements in the lattice vertex algebra $V_Q$:
\[
e_{k l}=\ep_{l k}e^{v_k-v_l}, \quad k\ne l.
\]
Let
\[
\mathring\fg=\bigoplus_{i=1}^{\ell-1}\C(v_i-v_{i+1})\oplus \bigoplus_{i\ne j}\C e_{i j}\subset V_Q.
\]
Note that $v_i-v_{i+1}$ stands for the element $(v_i-v_{i+1})s^{-1}\otimes e^0$ in $V_Q$, as pointed out in the previous section.
\begin{prp}\label{thm-sl}
On $\mathring\fg$ there is a Lie bracket defined by
\[
[a,b]=a_{(0)}b, \quad a, b\in\mathring\fg.
\]
Moreover, there is an isomorphism of Lie algebras given by the linear map
\begin{equation}\label{rho}
\rho\ :\mathring{\fg}\to \fsl_\ell
\end{equation}
such that
\[
\rho(v_i-v_{i+1})=E_{i, i}-E_{i+1,i+1}, \quad  \rho(e_{k l})=E_{k, l},
\]
with $1\le i\le \ell-1$ and $k\ne l$. Here $E_{i, j}$ stands for the $\ell\times\ell$ matrix with the $(i,j)$-component being $1$ and others being zero.
\end{prp}
\begin{prf}
Suppose $i,j\in\{1,2,\dots,\ell-1\}$; $k,l,k',l'\in\{1,2,\dots,\ell\}$ such that $k\ne l$ and $k'\ne l'$. By using \eqref{hmm} and \eqref{ealz} it is straightforward to verify the following equalities:
\begin{align*}
 [v_i-v_{i+1}, v_{j}-v_{j+1}]=&0,
 \\
 [v_i-v_{i+1}, e_{k l}]=&\left(\de_{i,k}- \de_{i+1,k}-\de_{i,l}+\de_{i+1,l}\right) e_{k l},
\\
[e_{k l}, v_i-v_{i+1}]=&  -[v_i-v_{i+1}, e_{k l}],
\end{align*}
and
\begin{align}
[e_{k l}, e_{k' l'}]=& \ep_{l k}\ep_{l' k'} \res_z E^{v_k-v_l}(z) \left(1\otimes e^{v_{k'}-v_{l'}}\right) \nonumber \\
=& \ep_{l k}\ep_{l' k'}\ve(v_k-v_l, v_{k'}-v_{l'})\cdot \nonumber \\
&\cdot\res_z z^{(v_k-v_l| v_{k'}-v_{l'})} \exp\left(\sum_{d>0}\frac{z^d}{d}(v_k-v_l)_{(-d)}\right)\otimes e^{v_k-v_l+v_{k'}-v_{l'}} \nonumber \\
=& \frac{\ep_{l k}\ep_{l' k'} \ep_{k k'}\ep_{l l'}}{\ep_{k l'}\ep_{l k'}} \cdot \nonumber \\
 &\cdot \res_z z^{\de_{k, k'}-\de_{k, l'}-\de_{l, k'}+\de_{l, l'} }
\left( 1+z(v_k-v_l)_{(-1)}\right)1\otimes e^{v_k-v_l+v_{k'}-v_{l'}}  \nonumber \\
=&\left\{
\begin{array}{cl}
 v_{l'}-v_{k'},  &  k=l',\ l=k'; \\
  e_{k l'},  &  k\ne l',\ l=k'; \\
 -e_{k' l},  & k=l', \ l\ne k'; \\
 0,  &  \hbox{ else}.
\end{array}
\right. \label{eekl}
\end{align}
The above equalities confirm that the map in \eqref{rho} is an isomorphism of Lie algebras. The proposition is proved.
\end{prf}

In the proof and in what follows, by the notation ``$\res$'' it means that $\res_z\left( \sum_p g_p z^p\right)=g_{-1}$.

According to the above proposition, there is a nondegenerate invariant bilinear form on $\mathring\fg$ given by
\[
(a|b)=\tr \left( \rho(a)\rho(b)\right), \quad a, b\in\mathring\fg.
\]
It is easy to verify
\begin{equation}\label{a1b}
a_{(1)}b=(a|b)\vac, \quad a,b\in\mathring\fg.
\end{equation}
One observes that, the bilinear form between elements like $v_i-v_j$ in the Lie algebra $\mathring\fg$ coincides with the bilinear form between them in the lattice $Q$. It is why we use the same notation $(\cdot|\cdot)$.

The following proposition can be verified with the help of  \eqref{ambnbra} and \eqref{a1b}.
\begin{prp}\label{thm-Lsl}
For any $a, b\in\mathring\fg\subset V_Q$ and $m,n\in\Z$, it holds that
\begin{equation}
\left[a_{(m)}, b_{(n)}\right]=[a,b]_{(m+n)}+\de_{m,-n}m(a|b)\mathrm{Id}_{V_Q}.
\end{equation}
\end{prp}
This proposition implies that
\begin{equation}\label{gaff}
\fg= \bigoplus_{m\in\Z}\bigoplus_{a\in\mathring\fg}\C a_{(m)}\oplus \C\mathrm{Id}_{V_Q}
\end{equation}
is isomorphic to the affine Lie algebra
\[
\mathcal{L}(\fsl_\ell)=\fsl_\ell[s, s^{-1}]\oplus \C K,
\]
of which the Lie bracket reads
\begin{equation}\label{aff}
[X s^m + \lambda K, Y s^n +\mu K] = [X, Y]s^{m+n}+\de_{m,-n}m\,\tr(X Y)K,
\end{equation}
with $X,Y\in\fsl_\ell$; $m,n\in\Z$ and $\lambda,\mu\in\C$.


\begin{rmk}
Since every semisimple Lie algebra has a faithful trace-free matrix realization (see \cite{Kac1990} for example), then Propositions~\ref{thm-sl} and \ref{thm-Lsl} imply that one can realize the semisimple Lie algebra as well as its affinization via a certain lattice vertex algebra.
\end{rmk}

\subsection{Vacuum orbit of the Lie group of $\mathcal{L}(\fsl_\ell)$}

Starting from the above lattice vertex algebra realization of the affine Lie algebra $\mathcal{L}(\fsl_\ell)$, we proceed to
derive the corresponding integrable hierarchy, in the from of Hirota equations.

It is known that the tensor product $V_Q\btimes V_Q$ is the space of states of the lattice vertex algebra associated to $Q\oplus Q$. More precisely, for this lattice vertex algebra, the vacuum vector is $\vac\btimes\vac$ and the state-field correspondence is given by
\begin{equation}\label{YY}
\mathds{Y}(a\btimes b,z)=\mathrm{Y}(a,z)\btimes \mathrm{Y}(b,z),\quad  a,b\in V_Q.
\end{equation}
It is easy to see that
\[
\mathds{Y}(a\btimes b,z)=\sum_{m,n\in \mathbb{Z}}a_{(m)}\btimes b_{(n)}z^{-m-n-2},\quad  a,b\in V_Q.
\]

Let us consider the following element
\begin{equation}\label{}
\Omega=\sum_{k=1}^\ell e^{v_k}\btimes e^{-v_k}\in V_Q\btimes V_Q,
\end{equation}
and its field
\begin{equation}\label{}
\mathds{Y}(\Omega,z)=\sum_{m\in\Z}\Omega_{(m)}z^{-m-1}.
\end{equation}

Recall the the symmetric algebra $S$ over the space $\fh^{<0}=\sum_{j<0}\fh s^{j}$. The completion of $S$ is denoted by
$\bar{S}$, whose topology is induced by the degree of the parameter $s$. We introduce
\begin{equation}\label{Vbar}
\bar{V}_Q=\bar{S}\otimes \C_\ve[Q].
\end{equation}
Clearly, every operator in ${\rm End}(V_Q)$ acts naturally on $\bar{V}_Q$.
\begin{thm}\label{thm-OmZZ}
Suppose that $X\in\fg$ satisfies $Z=e^X\vac\in \bar{V}_Q$, then the following equalities hold:
\begin{equation}\label{OmZZ}
\Omega_{(m)}\left(Z\btimes Z\right)=0, \quad m\ge0.
\end{equation}
\end{thm}

\begin{prf}
Thanks to \eqref{amvac}, one has
\[
\Omega_{(m)}\left(|0\rangle\btimes|0\rangle\right)=0, \quad m\geq0.
\]
In order to prove \eqref{OmZZ}, it suffices to show that, for $m\ge0$ and $X\in\fg$,
\[
\left[ \Om_{(m)}, e^X \btimes e^X \right]=0,
\]
which is equivalent to
\begin{equation}\label{OmX}
\left[ \Om_{(m)}, X \btimes \id_{\bar{V}_Q} +\id_{\bar{V}_Q}\btimes X \right]=0.
\end{equation}
To this end, one only needs to verify the equality \eqref{OmX} for $X=a_{(n)}$ with arbitrary
\[
a\in \left\{ v_i-v_{i+1}, e_{k l} \mid 1\le i\le\ell-1, \ k\ne l \right\}
\]
and $n\in\Z$.
In fact, for $t\ge0$ one has
\begin{align*}
&((v_i-v_{i+1})\btimes\vac+\vac\btimes(v_i-v_{i+1}) )_{(t)}\Omega \\
=&\sum_{j=1}^\ell \left( (v_i-v_{i+1})_{(t)} e^{v_j}\btimes e^{-v_j}+e^{v_j}\btimes(v_i-v_{i+1})_{(t)}e^{-v_j} \right) \\
=&\de_{t,0}\sum_{j=1}^\ell \left( (v_i-v_{i+1}|v_j) e^{v_j}\btimes e^{-v_j}+ (v_i-v_{i+1}|-v_j) e^{v_j}\btimes e^{-v_j} \right) \\
=& 0, \\
&(e_{k l}\btimes\vac+\vac\btimes e_{k l} )_{(t)}\Omega \\
=&\ep_{l k}\sum_{j=1}^\ell \res_z z^t \left( E^{v_k-v_l}(z) e^{v_j}\btimes e^{-v_j}+e^{v_j}\btimes E^{v_k-v_l}(z) e^{-v_j} \right) \\
=&\ep_{l k} \de_{t,0} \sum_{j=1}^\ell \left( \de_{l, j} e^{v_k-v_l}e^{v_j}\btimes e^{-v_j}+ \de_{k,j} e^{v_j}\btimes e^{v_k-v_l}e^{-v_j} \right)  \\
=& \ep_{l k} \de_{t,0}\left( \ve(v_k-v_l,v_l) e^{v_k}\btimes e^{-v_l} +  \ve(v_k-v_l,-v_k) e^{v_k}\btimes e^{-v_l} \right) \\
=& \ep_{l k} \de_{t,0}\left( -\ve(v_k,v_l) - \ve(v_l,v_k)\right) e^{v_k}\btimes e^{-v_l} \\
=& 0,
\end{align*}
hence
\begin{align*}
& \left[ a_{(n)} \btimes \id_{\bar{V}_Q} +\id_{\bar{V}_Q}\btimes a_{(n)}, \Om_{(m)}  \right] \\
= &
\left[ ( a\btimes\vac+\vac\btimes a)_{(n)}, \Om_{(m)}  \right]  \\
= & \sum_{t\ge0} \binom{n}{t}\left( ( a\btimes\vac+\vac\btimes a)_{(t)}\Om \right)_{(n+m-t)} =0,
\end{align*}
which confirms \eqref{OmX}. Therefore the theorem is proved.
\end{prf}

\subsection{Hirota bilinear equations associated to $\mathcal{L}(\fsl_\ell)$}
\label{sec-Hirsl}

We proceed to rewrite \eqref{OmZZ} in the form of Hirota bilinear equations. Consider the variables $\bs{x}=(\bs{x}^{(1)},\dots,\bs{x}^{(\ell)})$
with  $\bs{x^{(i)}}=\left(x_1^{(i)},x_2^{(i)},x_3^{(i)},\dots\right)$ and $\bs{u}=(u_1,\dots,u_\ell)$, which are assumed to commute with each other.
Let
\[
\mathcal{B}=\C[[\bs{x}]][\bs{u},\bs{u}^{-1}],
\]
where $\bs{u}^{-1}=\left( u_1^{-1},\dots,u_\ell^{-1}\right)$. Define a linear map
\begin{equation}\label{PiVQ}
\Pi\ : \ \bar{V}_Q\rightarrow \mathcal{B}
\end{equation}
by $\Pi(\vac)=1$ and
\begin{align*}
&\Pi\left( v_{i_1}s^{-m_1} \dots v_{i_k}s^{-m_k}\otimes e^{j_1 v_1+\dots +j_\ell v_\ell}\right)
=m_1\dots m_k x_{m_1}^{(i_1)}\dots x_{m_k}^{(i_k)} u_1^{j_1}\dots u_\ell^{j_\ell}
\end{align*}
for any $m_j\in\Z_{>0}$ and $j_i\in\Z$. Clearly, the map $\Pi$ is an isomorphism of linear spaces.

Given a vector $\al =\sum_{i=1}^\ell j_i v_i\in Q$, we write its coordinate with respect to the basis $(v_1,v_2,\dots,v_\ell)$ as $\underline{\al}=(j_1,j_2,\dots,j_\ell)\in\Z^\ell$, and for convenience
\[
\bs{u}^{\underline{\al} }=u_1^{j_1}\dots u_\ell^{j_\ell}.
\]
In particular, one has
\[
\bs{u}^{\underline{\al +v_i} }=u_1^{j_1}\dots u_{i-1}^{j_{i-1}}u_{i}^{j_i+1}u_{i+1}^{j_{i+1}} u_\ell^{j_\ell}.
\]
It is easy to see the following lemma.
\begin{lem}\label{thm-Pi}
Suppose that $f(\bs{x})\in\C[[\bs{x}]]$ and $\al \in Q$. For $1\le i\le\ell$ and $m\in\Z_{>0}$, the following equalities hold:
\begin{align} \label{Pivm}
\Pi\circ (v_{i})_{(m)}\circ \Pi^{-1}(f(\bs{x})\bs{u}^{\underline{\al} })= & \frac
{\pa f(\bs{x})}{\partial {x_{m}^{(i)}}}\bs{u}^{\underline{\al} },\\
\Pi\circ (v_{i})_{(-m)} \circ \Pi^{-1}(f(\bs{x})\bs{u}^{\underline{\al} })= & m  x_{m}^{(i)}f(\bs{x})\bs{u}^{\underline{\al} }, \\
\Pi\circ (v_{i})_{(0)} \circ \Pi^{-1}\left(f(\bs{x})\bs{u}^{\underline{\al} }\right)=&(v_{i}|\al )f(\bs{x})\bs{u}^{\underline{\al} }, 
\end{align}
and
\begin{align}
&\Pi\circ\mathrm{Y}(e^{\pm v_i},z)\circ\Pi^{-1}\left(f(\bs{x})\bs{u}^{\underline{\al} }\right) \nonumber \\
= & \varepsilon(\pm v_i,\al )z^{\pm (v_i|\al )}e^{\pm\xi(x^{(i)},z)}f\left(\bs{x}\mp [z^{-1}]^{(i)}\right)\bs{u}^{\underline{\al \pm v_i} }, \label{PiYe}
\end{align}
where
\[
\xi(\bs{x}^{(i)},z)=\sum\limits_{n>0}x^{(i)}_n z^n, \quad \left(\bs{x}\mp [z^{-1}]^{(i)}\right)=\left.\bs{x}\right|_{\bs{x}^{(i)}\mapsto\bs{x}^{(i)}\mp[z^{-1}]},
\]
with
\[
 [z^{-1}]=
\left(\frac{1}{z},\frac{1}{2 z^2},\frac{1}{ 3 z^{3}},\dots\right).
\]
\end{lem}

For $X\in\fg$ given in Theorem~\ref{thm-OmZZ}, it is called
\[
\tau(\bs{x},\bs{u})=\Pi(e^X\vac),
\]
the tau function for the vacuum orbit of the Lie group of $\fg\cong\mathcal{L}(\fsl_{\ell})$.
Clearly, it can be expanded in the form
\begin{equation}\label{tauQ}
\tau(\bs{x},\bs{u})=\sum_{\al\in  Q_0 } \tau_\al(\bs{x})\bs{u}^{\underline{\al}},
\end{equation}
where $ Q_0 $ is the root lattice of $\mathring\fg\cong\fsl_\ell$, say,
\begin{equation}\label{Q0}
 Q_0 =\left\{ \al=\sum_{i=1}^\ell j_i v_i \mid j_i\in\Z,\, \sum_{i=1}^\ell j_i=0 \right\}.
\end{equation}
In what follows, we will also call $\tau_\al(\bs{x})$ tau functions associated to the affine Lie algebra $\fg$.
\begin{thm}\label{thm-bl}
The tau functions associated to $\fg$ satisfy the following Hirota bilinear equations:
\begin{align}
\sum_{k=1}^{\ell} &\varepsilon(v_k,\al-\beta+v_i+v_j) {\rm Res}_z  \bigg( z ^{(v_k|\al-\beta)+m-2+\de_{i, k}+\de_{j, k}} e^{\xi\left(\bs{x}^{(k)}-\bs{x'}^{(k)},z\right)} \cdot \nonumber\\
& \cdot
\tau_{\al+v_i-v_k}\left(\bs{x}-[z^{-1}]^{(k)}\right) \tau_{\beta-v_j +v_k}\left(\bs{x}'+[z^{-1}]^{(k)}\right)\bigg)=0,
\label{bletauij}
\end{align}
with arbitrary $m\in\Z_{\geq0}$, $\al,\beta\in Q_0 $ and $i,j\in\{1,2,\dots,\ell\}$.
\end{thm}
\begin{prf}
Given any $Z=e^X\vac$ with $X\in\fg$, one writes
\[
\Pi(Z)\btimes\Pi(Z)=\sum_{\al,\beta\in Q_0 } \tau_\al(\bs{x})\tau_\beta(\bs{x'})\bs{u}^{\underline{\al}}\bs{u'}^{\underline{\beta}}.
\]
For $m\ge0$, according to Theorem~\ref{thm-OmZZ}, the equality \eqref{YY} and Lemma~\ref{thm-Pi}, one has
\begin{align*}
0=& \res_z z^m (\Pi\btimes\Pi) \circ \mathds{Y}(\Om,z)\circ (\Pi^{-1}\btimes\Pi^{-1}) \left( \Pi(Z)\btimes\Pi(Z) \right) \\
=& \sum_{k=1}^\ell \res_z z^m (\Pi \circ \mathrm{Y}(e^{v_k},z)\circ \Pi^{-1})\btimes (\Pi \circ \mathrm{Y}(e^{-v_k},z)\circ \Pi^{-1})  \left( \Pi(Z)\btimes\Pi(Z) \right) \\
=& \sum_{k=1}^\ell\sum_{\al,\beta\in Q_0 } \res_z \bigg( z^m \ve(v_k,\al)\ve(-v_k,\beta) z^{(v_k|\al)+(-v_k|\beta)} e^{\xi\left(\bs{x}^{(k)}-\bs{x'}^{(k)},z\right)} \cdot \\
&\qquad \cdot \tau_{\al}\left(\bs{x}-[z^{-1}]^{(k)}\right)\tau_{\beta}\left(\bs{x}'+[z^{-1}]^{(k)}\right)
\bs{u}^{\underline{\al+v_k}}\bs{u'}^{\underline{\beta-v_k} } \bigg) \\
=& \sum_{k=1}^\ell\sum_{\al^\prime,-\beta^\prime\in {Q}_1} \res_z \bigg( z^m \ve(v_k,\al^\prime-v_k)\ve(-v_k,\beta^\prime+v_k) z^{(v_k|\al^\prime-v_k)+(-v_k|\beta^\prime+v_k)} \cdot \\
& \cdot e^{\xi\left(\bs{x}^{(k)}-\bs{x'}^{(k)},z\right)}\tau_{\al^\prime - v_k} \left(\bs{x}-[z^{-1}]^{(k)}\right)
\tau_{\beta^\prime+v_k}\left(\bs{x}'+[z^{-1}]^{(k)}\right)
\bs{u}^{\underline{\al^\prime}}\bs{u'}^{\underline{\beta^\prime} } \bigg)\\
=& \sum_{k=1}^\ell\sum_{\al^\prime,-\beta^\prime\in {Q}_1} \res_z \bigg( \ve(v_k,\al^\prime)\ve(v_k,-\beta^\prime) z^{(v_k|\al^\prime-\beta^\prime)+m-2}  e^{\xi\left(\bs{x}^{(k)}-\bs{x'}^{(k)},z\right)} \cdot \\
& \cdot \tau_{\al^\prime - v_k}\left(\bs{x}-[z^{-1}]^{(k)}\right)
\tau_{\beta^\prime+v_k}\left(\bs{x}'+[z^{-1}]^{(k)}\right)
\bs{u}^{\underline{\al^\prime}}\bs{u'}^{\underline{\beta^\prime }} \bigg),
\end{align*}
where (cf.\eqref{Q0})
\begin{equation}\label{Q1}
Q_1=\left\{ \sum_i^\ell j_i v_i\mid j_i\in\Z,\, \sum_{i=1}^\ell j_i=1\right\}.
\end{equation}
One takes the coefficients of $\bs{u}^{\underline{\al^\prime} }\bs{u'}^{\underline{\beta^\prime}}$, and does the replacements
\[
\al^\prime\mapsto \al +v_i, \quad \beta^\prime\mapsto \beta-v_j,
\]
then concludes the theorem.
\end{prf}

\begin{rmk}
Since $\al$ and $\beta$ run over the root lattice $ Q_0 $ in \eqref{bletauij} and $\ve(v_k,-\beta+v_j)=\ve(v_k,\beta-v_j)$, then by taking $i=j=\ell$ and the fact $\ve(v_k,2 v_\ell)=\ve(v_k,v_\ell)^2=1$,
the bilinear equation \eqref{bletauij} can be written in the form
\begin{align}
\sum_{k=1}^{\ell} &\varepsilon(v_k,\al+\beta) {\rm Res}_z  \bigg( z ^{(v_k|\al-\beta)+m-2+2\de_{k,\ell}} e^{\xi\left(\bs{x}^{(k)}-\bs{x'}^{(k)},z\right)} \cdot \nonumber\\
& \cdot
\tau_{\al+v_\ell-v_k}\left(\bs{x}-[z^{-1}]^{(k)}\right)\tau_{\beta-v_\ell +v_k}\left(\bs{x}'+[z^{-1}]^{(k)}\right)\bigg)=0.
\label{bletauij2}
\end{align}
This equation coincides with equation~(1.69) obtained in \cite{KvdL-KPred} by taking all $\lambda_j=1$. In other words, the bilinear equation \eqref{bletauij2} is called the $(1,1,\dots,1)$-reduction of the $\ell$-component KP hierarchy (cf. \cite{Jimbo1983, KvdL-nKP}), which was derived  by Kac and van de Leur from a multicomponent boson-fermion correspondence.
\end{rmk}

\subsection{Lax equations  associated to $\mathcal{L}(\fsl_\ell)$} \label{sec-Laxsl}
The bilinear equation \eqref{bletauij} can be recast to the form of Lax equations. To this end, let us introduce a derivation
\[
\pa=\sum_{i=1}^\ell \frac{\pa}{\pa x_1^{(i)} },
\]
and the following matrix pseudo-differential operators
\begin{equation}\label{Pal}
P_\al=\sum_{n\ge0}P_{\al,n} \pa^{-n}
\end{equation}
with $\al\in Q_0 $ and $P_{\al,n}$ being $\ell\times\ell$ matrices whose $(i,j)$-component reads
\begin{equation}\label{Pnij}
\left(P_{\al,n}\right)_{i j}=-\ve(v_j,v_i)\res_z \frac{z^{n+\de_{i, j}-2}\tau_{\al+v_i-v_j}\left(\bs{x}-[z^{-1}]^{(j)}\right)}{\tau_{\al}(\bs{x})}.
\end{equation}
In particular, the leading term of $P_\al$ is just the identity matrix, say,
\[
P_{\al,0}=-\mathrm{diag}\left(\ve(v_1,v_1),\ve(v_2,v_2),\dots,\ve(v_\ell,v_\ell) \right)=I.
\]
Hence the operators $P_\al$ are invertible, namely,
\[
{P_\al}^{-1}=I+\sum_{k\ge1}(I-P_\al)^k.
\]

Given a pseudo-differential operator $f\pa^n$ with $f$ being a scalar function, its adjoint operator means
\[
(f\pa^n)^*=(-\pa)^n f;
\]
when $n\ge0$ one has the actions
\begin{equation}\label{panexp}
f \pa^n e^{\xi(\bs{x}^{(i)},z)}=f z^n e^{\xi(\bs{x}^{(i)},z)}, \quad 1\le i\le \ell,
\end{equation}
and when $n<0$ it is assumed the equalities \eqref{panexp} by convention.
One introduces the following diagonal matrices
\begin{align*}
\Xi(\bs{x},z)=& \mathrm{diag}\left(\xi(\bs{x}^{(1)},z),\xi(\bs{x}^{(2)},z),\dots,\xi(\bs{x}^{(\ell)},z)  \right), \\
\Lambda_\al (z)=& \mathrm{diag}\left(\ve(v_1,\al)z^{(v_1|\al)},  \ve(v_2,\al)z^{(v_2|\al)},\dots,\ve(v_\ell,\al)z^{(v_\ell|\al)} \right)
\end{align*}
with $\al\in Q_0 $. In particular, one has $\Lambda_0(z)=I$. With a similar method as in \cite{Jimbo1983}, we obtain the following result.
\begin{thm} \label{thm-Laxbl}
The bilinear equations \eqref{bletauij} are equivalent to
\begin{equation}\label{blPexp}
\res_z \left( z^m \left(P_\al e^{\Xi(\bs{x},z)}\right)\cdot\Lambda_{\al-\beta}(z)\cdot \left.\left( (P_\beta^*)^{-1} e^{-\Xi(\bs{x},z)} \right)^T\right|_{\bs{x}\mapsto\bs{x}'}\right)=0,
\end{equation}
with $m\in \Z_{\ge0}$, $\al,\beta\in Q_0 $ and arbitrary variables $\bs{x}$ and $\bs{x}^{\prime}$. Here the superscript ``$T$'' stands for the transpose of matrices. Moreover, the operators $P_\al$ satisfy
\begin{align}\label{PmPinv}
 & \left(P_\al \pa {P_{\al}}^{-1}\right)_{-}=0, \\
 & \frac{\pa P_\al}{\pa x^{(i)}_k}=-\left(P_\al E_{i i} \pa^k {P_{\al}}^{-1}\right)_{-}P_\al, \quad 1\leq i\leq\ell,\label{Pxik}
\end{align}
which give the Lax equations of the $(1,1,\dots,1)$-reduction of the $\ell$-component KP hierarchy  (see in \cite{KvdL-KPred}). Note that the subscript ``$-$'' means to take the negative powers in $\pa$.
\end{thm}
\begin{prf}
Let us sketch the proof that is similar to the case of the KP hierarchy. Firstly, we rewrite the bilinear equations \eqref{bletauij} into the form
\[
\res_z \left( z^m \left(P_\al e^{\Xi(\bs{x},z)}\right)\cdot\Lambda_{\al-\beta}(z) \cdot\left.\left( \tilde{P}_\beta  e^{-\Xi(\bs{x},z)} \right)^T\right|_{\bs{x}\mapsto\bs{x}'}\right)=0,
\]
where $\tilde{P}_{\beta}$ are certain matrix pseudo-differential operators with leading term being the identity matrix. One lets $\pa^n$ with $n\ge0$ act on the above equation and
takes $\beta=\al$ and $\bs{x}'=\bs{x}$, then obtains
\[
 \left(\pa^n P_\al \pa^m \tilde{P}_{\al}^* \right)_{-}=0, \quad m\ge0,
\]
which implies $\tilde{P}_\al=(P_\al^*)^{-1}$ and hence \eqref{blPexp} and \eqref{PmPinv}.
Taking $\beta=\al$ in \eqref{blPexp}, and let $\pa^n\frac{\pa}{\pa x^{(i)}_k}$ with $n\ge0$ act on it, then one derives the equations \eqref{Pxik}. The theorem is proved.
\end{prf}

This theorem yields the following result.
\begin{cor}
For any $\al\in Q_0$ and $k\ge0$, it holds that
\begin{equation}\label{paPal}
\sum_{i=1}^\ell\frac{\pa P_\al}{\pa x^{(i)}_k}=0.
\end{equation}
\end{cor}
Observe that the coefficients of $(P_{\al,n})_{j j}$ in \eqref{Pnij} are differential polynomials in $\log\tau_\al$. In combination with \eqref{paPal}, it implies that
\begin{equation}\label{patau}
\frac{\pa }{\pa x_n^{(j)} }\sum_{i=1}^\ell\frac{\pa \log\tau_\al}{\pa x^{(i)}_k}=0
\end{equation}
for any $1\le j\le\ell$ and $n,k\ge1$,
hence $\sum_{i=1}^\ell\frac{\pa \log\tau_\al}{\pa x^{(i)}_k}$ are constants. In the sequence, such constants are assumed to be zero (cf. \cite{Jimbo1983}), due to the freedom of solutions of the bilinear equation \eqref{bletauij}.

\section{Realization of $\mathcal{L}_{r+1}^\tor( \fsl_\ell)$ and the associated integrable hierarchy}\label{untoridial}

In this section, let us fix a positive integer $r$, and denote by $\mathcal{L}_{r+1}^\tor( \fsl_\ell)$ the toroidal Lie algebra corresponding to $\fsl_\ell$ with $r+1$ loop parameters (see Appenix~A for the definition). In a fairly straightforward way, we will write down a realization of $\mathcal{L}_{r+1}^\tor(\fsl_\ell)$ in the framework of lattice vertex algebras \cite{Bakalov2004,Bakalov2021}, and then derive an integrable hierarchy of bilinear equations of tau functions associated to $\mathcal{L}_{r+1}^\tor( \fsl_\ell)$.

\subsection{From lattice vertex algebras to torodial Lie algebras }\label{untwistedtorol}

Recall the lattice $Q=\oplus_{i=1}^\ell \Z v_i$ of rank $\ell$ in the previous section. Following \cite{Bakalov2021,ERM}, one can extend this lattice to $\Ga=Q\oplus R$, where $R$ is a lattice of rank $2 r$ as
\[
R=\bigoplus_{\nu=1}^{r}\left( \Z\ta_\nu\oplus\Z \Ta_\nu\right).
\]
The nondegenerate symmetric bilinear form on $Q$ is extended to $\Ga$ by
\begin{equation}\label{Gabl}
(\ta_\nu|\Ta_\mu)=\de_{\nu,\mu}, \quad (v_i|\ta_\nu)=(v_i|\Ta_\mu)=(\ta_\nu|\ta_\mu)=(\Ta_\nu|\Ta_\mu)=0,
\end{equation}
with $1\le\nu,\mu\le r$ and $1\le i\le\ell$. Meanwhile the bi-multiplicative function on $Q\times Q$  is extended to
\[
\ve:\Ga\times\Ga\to \{\pm 1\}
\]
by
\begin{equation}\label{Gaep}
\varepsilon\left(\ta_\nu,\Ta_\mu\right)=(-1)^{\de_{\nu,\mu}},\quad \ve(\ta_\nu,\ta_\mu)=\ve(\Ta_\nu,\Ta_\mu)=1,
\end{equation}
with  $1\le\nu,\mu\le r$, and
\begin{equation}\label{QGaep}
\ve(v,\eta)=\ve(\eta,v)=1, \quad \forall v\in Q, \ \eta\in R.
\end{equation}

In the same way as before, one has a lattice vertex algebra associated to $\Ga$, for which the state space is $V_\Ga$, the vacuum vector is  denoted by $\bs { \vac}$ and the state-field correspondence denoted by $\mathrm{Y}(\cdot,z)$ by abusing notations.

For convenience we write $\bs{\ta}=(\ta_1,\ta_2,\dots,\ta_r)$, and
\[
\bs{p}\cdot\bs{\ta}=\sum_{\nu=1}^r p_\nu\ta_\nu\in R\subset \Ga
\]
for  $\bs{p}=(p_1,p_2,\dots,p_r)\in\Z^r$.
In the lattice vertex algebra associated to $\Ga$, let us consider the following elements:
\begin{align}\label{hpm}
h_{i,m}^{\bs{p}}=&\left( (v_i-v_{i+1})s^{-1}\otimes e^{\bs{p}\cdot\bs{\ta}}\right)_{(m)}, \\
e_{k l,m}^{\bs{p}}=&\ep_{l k}\left( 1\otimes e^{v_k-v_l+\bs{p}\cdot\bs{\ta}}\right)_{(m)},  \label{epm}
\\
\ka_{\nu,m}^{\bs{p}}=&\left\{ \begin{array}{cl}
                              \left( 1\otimes e^{\bs{p}\cdot\bs{\ta}}\right)_{(m-1)}, & \nu=0; \\
                             \left( \ta_\nu s^{-1}\otimes e^{\bs{p}\cdot\bs{\ta}}\right)_{(m)}, & 1\le\nu\le r,
                            \end{array}
                            \right. \label{kapm}
\end{align}
where $1\le i\le \ell-1$, $m\in\Z$, $\bs{p}\in\Z^r$, and  $ k, l\in\{1,2,\dots,\ell\}$ being distinct.

\begin{prp}\label{thm-hep}
The elements \eqref{hpm}--\eqref{kapm} satisfy the following equalities:
\begin{align}
\left[ h_{i,m}^{\bs{p}},h_{j,n}^{\bs{q}}\right]=& (2\de_{i,j}-\de_{i,j+1}-\de_{i+1,j})\left( m \ka_{0,m+n}^{\bs{p}+\bs{q}}+\sum_{\nu=1}^r p_\nu \ka_{\nu,m+n}^{\bs{p}+\bs{q}} \right),  \label{hhpm}
\\
\left[ h_{i,m}^{\bs{p}},e_{k l,n}^{\bs{q}}\right]=& (\de_{i, k}+\de_{i+1,l}-\de_{i, l}-\de_{i+1,k}) e_{k l,m+n}^{\bs{p}+\bs{q}}, \label{hepm}
\\
\left[ e_{k l,m}^{\bs{p}},e_{k' l',n}^{\bs{q}}\right]=&\left\{
\begin{array}{cl}
 \left((v_k-v_l)s^{-1}\otimes e^{(\bs{p}+\bs{q})\cdot\bs{\ta} }\right)_{(m+n)} & \\
 \quad + m \ka_{0,m+n}^{\bs{p}+\bs{q}}
  +\sum_{\nu=1}^r p_\nu \ka_{\nu,m+n}^{\bs{p}+\bs{q}} ,  &  k=l',\ l=k'; \\
  e_{k l', m+n}^{\bs{p}+\bs{q}},  &  k\ne l',\ l=k'; \\
 -e_{k' l, m+n}^{\bs{p}+\bs{q}},  & k=l', \ l\ne k'; \\
 0,  &  \hbox{ else},
\end{array}
\right. \label{epep}
\\
\left[ \ka_{\nu,m}^{\bs{p}}, a\right]=&0. \label{kabra}
\end{align}
Here $m,n\in\Z$; $\bs{p},\bs{q}\in\Z^r$; $1\le i,j\le\ell-1$; $\nu,k,l,k',l'\in\{1,2,\dots,\ell\}$ such that $k\ne l$ and $k'\ne l'$, and $``a"$ stands for any element given in \eqref{hpm}--\eqref{kapm}.
\end{prp}

This proposition can be verified in a straightforward way, which will be given in Appendix~B.
Observe that the elements $\left((v_k-v_l)s^{-1}\otimes e^{(\bs{p}+\bs{q})\cdot\bs{\ta} }\right)_{(m+n)}$ in \eqref{epep} are just linear combinations of elements of the form $h_{i,m+n}^{\bs{p}+\bs{q}}$. Then according to Proposition~\ref{thm-hep}, the elements \eqref{hpm}--\eqref{kapm} indeed generate a Lie algebra, which is denoted by $\fg^\tor$. More exactly, it leads to the following result.
\begin{prp}
The Lie algebra $\fg^\tor$ is isomorphic to the toroidal Lie algebra $\mathcal{L}_{r+1}^\tor(\fsl_\ell)$ (see Appendix~A), via an isomorphism given by:
\begin{align*}
h_{i,m}^{\bs{p}}\mapsto & (E_{i, i}-E_{i+1,i+1})\otimes t_0^m t_1^{p_1}\dots t_r^{p_r}, \\
e_{k l,m}^{\bs{p}} \mapsto & E_{k, l}\otimes t_0^m t_1^{p_1}\dots t_r^{p_r}, \\
\ka_{\nu,m}^{\bs{p}}\mapsto &
\overline{ t_0^m t_1^{p_1}\dots t_r^{p_r}K_\nu}.
\end{align*}
\end{prp}

\begin{rmk}
The above proposition is a reformulation of the type-A case of Theorem~3.14 in \cite{ERM} by Eswara Rao and Moody in the context of vertex operators. It can also be considered as a corollary of Theorem~4.2 in \cite{Bakalov2021} by Bakalov and Kirk, which is based on a general setting and even holds for an arbitrary simple Lie algebra with an automorphism of order $N$. From this point of view, Proposition~\ref{thm-hep} can be considered as a corollary of such results.
\end{rmk}

\subsection{Vacuum orbit of the Lie group of $\mathcal{L}^\tor_{r+1}(\fsl_\ell)$ }\label{ }

Now we are to derive a hierarchy of Hirota equations to characterize the vacuum orbit of the Lie group of the toroidal Lie algebra $\fg^\tor\cong\mathcal{L}^\tor_{r+1}(\fsl_\ell)$.

It is known that $V_\Ga\btimes V_\Ga$ is the state space of the lattice vertex algebra associated to $\Ga\btimes\Ga$, whose state-filed correspondence is denoted as $\mathbb{Y}(\cdot,z)$ (cf. \eqref{YY}). For $\bs{q}\in\Z^r$, we introduce
\begin{align}
\Omega^{{\rm tor}}_{\bs{q}}=\sum_{k=1}^{\ell} \left(e^{v_k+\bs{q}\cdot\bs{\ta}}\btimes  e^{-v_k-\bs{q}\cdot\bs{\ta}}\right)\in V_\Ga\btimes V_\Ga,
\end{align}
and consider its field
\begin{equation}\label{}
\mathbb{Y}(\Omega^\tor_{\bs{q}},z)=\sum_{m\in\Z} \left(\Om^\tor_{\bs{q}}\right)_{(m)}z^{-m-1}.
\end{equation}
Let
\[
\Om^\tor_{(m)}=\sum_{\bs{q}\in\Z^r}\left(\Om^\tor_{\bs{q}}\right)_{(m)}, \quad m\in\Z.
\]
According to \eqref{amvac}, we have
\begin{equation}\label{Omvactor}
\Om^\tor_{(m)}\left(\bs{\vac}\btimes\bs{\vac} \right)=0, \quad m{\ge0}.
\end{equation}

In the same way as \eqref{Vbar}, for the lattice vertex algebra $V_\Ga$ we consider its completion space $\bar{V}_\Ga$.
\begin{thm}\label{thm-OmZZtor}
Suppose that $X\in\fg^\tor$ satisfying $Z^\tor=e^X\bs{\vac}\in \bar{V}_\Ga$, then the following equalities hold:
\begin{equation}\label{OmZZ}
\Omega^\tor_{(m)}\left(Z^\tor\btimes Z^\tor\right)=0, \quad m\ge0.
\end{equation}
\end{thm}

\begin{prf}
The proof is similar with that of Theorem~\ref{thm-OmZZ}. Thanks to \eqref{Omvactor}, it suffices to show
\begin{equation}\label{OmXtor}
\left[ \Om^\tor_{(m)}, X \btimes \id_{\bar{V}_\Gamma} +\id_{\bar{V}_\Gamma}\btimes X \right]=0, \quad m\ge0,
\end{equation}
for $X=a_{(n)}$ with arbitrary $n\in\Z$ and
\[
a\in \left\{ (v_i-v_{i+1})s^{-1}\otimes e^{\bs{p}\cdot\bs{\ta}}, 1\otimes e^{v_k-v_l+\bs{p}\cdot\bs{\ta}} \mid 1\le i\le\ell-1; \ 
k\ne l;\ \bs{p}\in\Z^r \right\}.
\]
In fact, for $a=(v_i-v_{i+1})s^{-1}\otimes e^{\bs{p}\cdot\bs{\ta}}$ and $t\ge0$, by using \eqref{Gabl}--\eqref{QGaep} one has
\begin{align*}
& a_{(t)}\left(1\otimes e^{v_j+\bs{q}\cdot\bs{\ta}}\right)
\\
=& \res_z \Bigg( z^t :\sum_{l\in\Z} (v_i-v_{i+1})_{(-l)}z^{l-1}\cdot \\
&\quad \cdot e^{\bs{p}\cdot\bs{\ta}} z^{(\bs{p}\cdot\bs{\ta})_{(0)}}  \exp\left(\sum_{d>0}(\bs{p}\cdot\bs{\ta})_{(-d)}\frac{z^d}{d}\right) \exp\left(\sum_{d<0}(\bs{p}\cdot\bs{\ta})_{(-d)}\frac{z^d}{d}\right): \left(1\otimes e^{v_j+\bs{q}\cdot\bs{\ta}}\right) \Bigg)
\\
=& \ve(\bs{p}\cdot\bs{\ta},v_j+\bs{q}\cdot\bs{\ta}) \res_z \Bigg( z^t \sum_{l\ge 1} (v_i-v_{i+1})_{(-l)}z^{l-1+(\bs{p}\cdot\bs{\ta}|v_j+\bs{q}\cdot\bs{\ta}) }\cdot \\
&\quad \cdot\exp\left(\sum_{d>0}(\bs{p}\cdot\bs{\ta})_{(-d)}\frac{z^d}{d}\right) \left(1\otimes e^{v_j+(\bs{p}+\bs{q})\cdot\bs{\ta}}\right) \Bigg) \\
&\quad + \res_z \Bigg( z^t e^{\bs{p}\cdot\bs{\ta}} z^{(\bs{p}\cdot\bs{\ta})_{(0)}} \exp\left(\sum_{d>0}(\bs{p}\cdot\bs{\ta})_{(-d)}\frac{z^d}{d}\right) \exp\left(\sum_{d<0}(\bs{p}\cdot\bs{\ta})_{(-d)}\frac{z^d}{d}\right)\cdot \\
 &\quad \cdot \sum_{l\le0} (v_i-v_{i+1})_{(-l)}z^{l-1} \left(1\otimes e^{v_j+\bs{q}\cdot\bs{\ta}}\right) \Bigg)  \\
=& \res_z \Bigg( z^{t-1}(v_i-v_{i+1})_{(0)} \exp\left(\sum_{d>0}(\bs{p}\cdot\bs{\ta})_{(-d)}\frac{z^d}{d}\right) \left(1\otimes e^{v_j+(\bs{p}+\bs{q})\cdot\bs{\ta}}\right) \Bigg)
\\
=&\de_{t, 0}(v_i-v_{i+1}|v_j)\otimes e^{v_j+(\bs{p}+\bs{q})\cdot\bs{\ta}}.
\end{align*}
Similarly, one has
\[
a_{(t)}\left(1\otimes e^{-v_j-\bs{q}\cdot\bs{\ta}}\right)=-\de_{t, 0}(v_i-v_{i+1}|v_j)\otimes e^{-v_j-(\bs{q}-\bs{p})\cdot\bs{\ta}}.
\]
Thus by using \eqref{YY} and \eqref{ambnbra} we obtain
\begin{align*}
& \left[ a_{(n)} \btimes \id_{\bar{V}\Gamma} +\id_{\bar{V}\Gamma}\btimes a_{(n)}, \Om_{(m)}^\tor  \right] \\
= &
\left[ ( a\btimes\bs{\vac}+\bs{\vac}\btimes a)_{(n)}, \Om_{(m)}^\tor  \right]  \\
= & \sum_{t\ge0} \sum_{\bs{q}\in\Z^r}\binom{n}{t}\left( ( a\btimes\bs{\vac}+\bs{\vac}\btimes a)_{(t)}\Om_{\bs{q}}^\tor \right)_{(n+m-t)} \\
=& \sum_{t\ge0} \binom{n}{t} \sum_{j=1}^\ell\sum_{\bs{q}\in\Z^r}\Bigg( a_{(t)}\left(e^{v_j+\bs{q}\cdot\bs{\ta} }\right)\btimes  e^{-v_j-\bs{q}\cdot\bs{\ta}}  + e^{v_j+\bs{q}\cdot\bs{\ta}}\btimes a_{(t)}\left(e^{-v_j-\bs{q}\cdot\bs{\ta}}  \right) \Bigg)_{(n+m-t)} \\
=& \sum_{j=1}^\ell(v_i-v_{i+1}|v_j) \sum_{\bs{q}\in\Z^r}\big(e^{v_j+(\bs{p}+\bs{q})\cdot\bs{\ta}}\btimes e^{-v_j-\bs{q}\cdot\bs{\ta}}-e^{v_j+\bs{q}\cdot\bs{\ta}} \btimes e^{-v_j-(\bs{q}-\bs{p})\cdot\bs{\ta}}\big)_{(n+m)} \\
=& \sum_{j=1}^\ell(v_i-v_{i+1}|v_j) \sum_{\bs{q}\in\Z^r}\big(e^{v_j+(\bs{p}+\bs{q})\cdot\bs{\ta}}\btimes e^{-v_j-\bs{q}\cdot\bs{\ta}}-e^{v_j+(\bs{p}+\bs{q})\cdot\bs{\ta}}\btimes e^{-v_j-\bs{q}\cdot\bs{\ta}}\big)_{(n+m)}=0.
\end{align*}

Similarly, for $b= 1\otimes e^{v_k-v_l+\bs{p}\cdot\bs{\ta}} $ and $t\ge0$, one has
\begin{align*}
& b _{(t)}\left(1\otimes e^{v_j+\bs{q}\cdot\bs{\ta}}\right)
\\
=& \res_z \Bigg( z^t   e^{v_k-v_l+\bs{p}\cdot\bs{\ta}} z^{(v_k-v_l+\bs{p}\cdot\bs{\ta})_{(0)}} \exp\left(\sum_{d>0}(v_k-v_l+\bs{p}\cdot\bs{\ta})_{(-d)}\frac{z^d}{d}\right) \\
&\quad \exp\left(\sum_{d<0}(v_k-v_l+\bs{p}\cdot\bs{\ta})_{(-d)}\frac{z^d}{d}\right) \left(1\otimes e^{v_j+\bs{q}\cdot\bs{\ta}}\right) \Bigg)
\\
=& \ve(v_k-v_l+\bs{p}\cdot\bs{\ta},v_j+\bs{q}\cdot\bs{\ta}) \res_z \Bigg( z^{t+(v_k-v_l+\bs{p}\cdot\bs{\ta}|v_j+\bs{q}\cdot\bs{\ta}) } \\
&\quad \exp\left(\sum_{d>0}(v_k-v_l+\bs{p}\cdot\bs{\ta})_{(-d)}\frac{z^d}{d}\right) \left(1\otimes e^{v_k-v_l+v_j+(\bs{p}+\bs{q})\cdot\bs{\ta}}\right) \Bigg) \\
=& \de_{t, 0}\de_{j, l} \frac{\ep_{k l}}{\ep_{l l}}\otimes e^{v_k+(\bs{p}+\bs{q})\cdot\bs{\ta}} \\
=& -\de_{t, 0}\de_{j, l} \ep_{k l} \otimes e^{v_k+(\bs{p}+\bs{q})\cdot\bs{\ta}}.
\end{align*}
In the same way, one has
\[
\left(b \right)_{(t)}\left(1\otimes e^{-v_j-\bs{q}\cdot\bs{\ta}}\right) = - \de_{t, 0}\de_{j, k} \ep_{l k} \otimes e^{-v_l-(\bs{q}-\bs{p})\cdot\bs{\ta}}.
\]
Similar as before, we obtain
\begin{align*}
& \left[ b_{(n)} \btimes \id_{\bar{V}\Gamma} +\id_{\bar{V}\Gamma}\btimes b_{(n)}, \Om_{(m)}^\tor  \right] \\
=& \sum_{t\ge0} \binom{n}{t} \sum_{j=1}^\ell\sum_{\bs{q}\in\Z^r}\Bigg( b_{(t)}\left(e^{v_j+\bs{q}\cdot\bs{\ta} }\right)\btimes  e^{-v_j-\bs{q}\cdot\bs{\ta}} + e^{v_j+\bs{q}\cdot\bs{\ta}}\btimes b_{(t)}\left(e^{-v_j-\bs{q}\cdot\bs{\ta}}  \right) \Bigg)_{(n+m-t)} \\
=& \sum_{\bs{q}\in\Z^r} \Bigg( -  \ep_{k l} e^{v_k+(\bs{p}+\bs{q})\cdot\bs{\ta}}\btimes  e^{-v_l-\bs{q}\cdot\bs{\ta}} - e^{v_k+\bs{q}\cdot\bs{\ta}}\btimes   \ep_{l k}  e^{-v_l-(\bs{q}-\bs{p})\cdot\bs{\ta}} \Bigg)_{(n+m)}
\\
=& -(\ep_{k l}+\ep_{l k}) \sum_{\bs{q}\in\Z^r} e^{v_k+(\bs{p}+\bs{q})\cdot\bs{\ta}}\btimes  e^{-v_l-\bs{q}\cdot\bs{\ta}}=
0.
\end{align*}

 Therefore, the equality \eqref{OmXtor} is verified, so the theorem is proved.
\end{prf}

\subsection{Hirota bilinear equations associated to $\mathcal{L}^\tor_{r+1}(\fsl_\ell)$}\label{sec-bilineartor}
Let us consider a subspace $\bar{V}^\tor$ of $\bar{V}_\Ga$ that is spanned by elements of the form:
\begin{equation}\label{Vtor}
v_{i_1} s^{-m_1}\cdot\dots\cdot v_{i_k} s^{-m_k} \theta_{\nu_1}s^{-n_1}\cdot\dots\cdot \theta_{\nu_l}s^{-n_l}\otimes e^{\al+\bs{p}\cdot\bs{\ta}}
\end{equation}
with $1\le i_j\le \ell$; $1\le \nu_d\le r$; $m_j,n_d > 0$; $k,l\ge0$; $\al\in Q$ and $\bs{p}\in\Z^r$. Clearly, one sees that $\bar{V}_Q\subset \bar{V}^\tor$.

Recall the variables $\bs{x}$ and $\bs{u}$ in Section~\ref{sec-Hirsl}. Now we introduce variables $\bs{y}=(\bs{y}^{(1)},\dots,\bs{y}^{(r)})$
with  $\bs{y^{(\nu)}}=\left(y_1^{(\nu)},y_2^{(\nu)},y_3^{(\nu)},\dots\right)$, and  $\bs{w}=(w_1,\dots,w_r)$; for convenience we write $e^{\bs{w} }=(e^{w_1},\dots,e^{w_r})$. Let us consider a commutative ring  as follows
\[
\mathcal{B}^\tor=\C\left[\left[\bs{x},\bs{y}\right]\right][ e^{\bs{w}}, e^{-\bs{w}},\bs{u},\bs{u}^{-1}].
\]
The isomorphism $\Pi$ given in \eqref{PiVQ} can be extended to
\[
\Pi\ : \ \bar{V}^\tor\rightarrow \mathcal{B}^\tor,
\]
such that $\Pi(\bs{\vac})=1$ and
\begin{align*}
&\Pi\left(v_{i_1} s^{-m_1}\cdot\dots\cdot v_{i_k} s^{-m_k} \ta_{\nu_1}s^{-n_1}\cdot\dots\cdot \ta_{\nu_l}s^{-n_l}\otimes e^{j_1 v_1+\dots +j_\ell v_\ell+p_1\ta_1+\dots+p_r\ta_r}\right) \\
=&m_1\dots m_k n_1\dots n_j x_{m_1}^{(i_1)}\dots x_{m_k}^{(i_k)} y _{n_1}^{(\nu_1)}\dots y_{n_l}^{(\nu_l)} u_1^{j_1}\dots u_\ell^{j_\ell}e^{p_1 w_1}\dots e^{p_r w_r}.
\end{align*}
It is easy to see that, the equalities \eqref{Pivm}--\eqref{PiYe} still hold true in the present circumstance. Moreover, for $1\le\nu\le r$ and $n\in\Z$ one has
\begin{equation}\label{Pita}
\Pi\circ(\ta_{\nu})_{(-n)}\circ\Pi^{-1}=\left\{ \begin{array}{cc}
                                                  n y_n^{(\nu)}, & n>0; \\
                                                  0, & n\le0.
                                                \end{array}
\right.
\end{equation}

Assume $Z^\tor=e^X\bs{\vac}\in \bar{V}^\tor$ to be given as in Theorem~\ref{thm-OmZZtor}, then it has the following expansion (cf.\eqref{tauQ})
\begin{equation}\label{tauQtor}
\Pi(Z^\tor)=\sum_{\al\in  Q_0 } \tau_\al(\bs{x},\bs{y},\bs{w})\bs{u}^{\underline{\al}}.
\end{equation}
These coefficients $\tau_\al$ are called the tau functions for the vacuum orbit of the Lie group of the toroidal Lie algebra $\fg^\tor\cong\mathcal{L}^\tor_{r+1}(\fsl_\ell)$.

\begin{thm}\label{thm-bltor}
The tau functions given in \eqref{tauQtor} satisfy the following Hirota bilinear equations:
for any $\al,\beta\in Q_0$ and $i,j\in\{1,2,\dots,\ell\}$,
\begin{align}
&\sum_{k=1}^{\ell} \ve(v_k,\al-\beta+v_i+v_j) \res_z \bigg(  z^{(v_k|\al-\beta)+m-2+\de_{i,k}+\de_{j, k}}
\nonumber\\
&\quad    e^{\xi(\bs{x}^{(k)}-\bs{x}'^{(k)},z)}
\tau_{\al+v_i-v_k}\left(\bs{x}-[z^{-1}]^{(k)},\bs{y}+\bs{b},\bs{w}-\xi(\bs{b},z)\right) \nonumber \\
&\quad  \tau_{\beta-v_j+v_k}\left(\bs{x'}+[z^{-1}]^{(k)},\bs{y}-\bs{b},\bs{w}+\xi(\bs{b},z)\right)
\bigg)=0, \label{eq-bltor}
\end{align}
where $\bs{b}=(\bs{b}^{(1)},\dots,\bs{b}^{(r)})$
with  $\bs{b^{(\nu)}}=\left(b_1^{(\nu)},b_2^{(\nu)},b_3^{(\nu)},\dots\right)$ being sequences of constants, and $\xi(\bs{b},z)=\left(\xi(\bs{b}^{(1)},z),\dots,\xi(\bs{b}^{(r)},z)\right)$.
\end{thm}

In order to prove the above theorem, the following lemma is useful (cf. Lemma~4 in \cite{KIT}).
\begin{lem}[\cite{Billig1999,ISW}] \label{thm-KIT}
Let  $\mathcal{A}=\mathbb{C}[e^{\zeta},e^{-\zeta}] \otimes
\mathbb{C}[\bs{z}]$ with $\bs{z}=(z_1, z_2, \ldots)$. For $\ga$ belonging to a finite set  $\Upsilon$, let $D_\ga(\zeta) = \sum\limits_{n\geq 0}
\zeta^n D_{\ga,n}$, where $D_{\ga,n}$
are differential operators acting on $\mathcal{A}$ that may depend on
${ \pa/ \pa \zeta}$ but not on $\zeta$.
If for some $g_\ga(\zeta, \bs{z}) \in \mathcal{A}$, it satisfies that
\[
\sum_{\ga\in \Upsilon}\sum\limits_{q\in \mathbb{Z}} e^{q \zeta} D_{\ga}(q) g_\ga(\zeta, \bs{z}) = 0,
\]
then
\[
\left.\sum_{\ga\in \Upsilon}D_\ga\left(\omega- \frac{\pa}{\pa \zeta}  \right) g_\ga(\zeta,\bs{ z})
 \right|_{\zeta=0} = 0,
\]
where $\omega$ is a parameter.
\end{lem}

\begin{prfof}{Theorem~\ref{thm-bltor}}
Similar as in the proof of Theorem~\ref{thm-bl}, thanks to \eqref{tauQtor} let us write
\[
\Pi(Z^\tor)\btimes\Pi(Z^\tor)=\sum_{\al,\beta\in Q_0 } \tau_\al(\bs{x},\bs{y},\bs{w})\tau_\beta(\bs{x'},\bs{y'},\bs{w'}) \bs{u}^{\underline{\al}}\bs{u'}^{\underline{\beta}}.
\]
For $m\ge0$, according to Theorem~\ref{thm-OmZZtor}, the equalities \eqref{YY} and \eqref{Pita}, we have
\begin{align*}
0=& \res_z z^m (\Pi\btimes\Pi) \circ \mathds{Y}(\Om^\tor,z)\circ (\Pi^{-1}\btimes\Pi^{-1}) \left( \Pi(Z^\tor)\btimes\Pi(Z^\tor) \right)
\\
=& \sum_{k=1}^\ell \sum_{\bs{q}\in\Z^r} \res_z\Big( z^m (\Pi \circ \mathrm{Y}(e^{v_k+\bs{q}\cdot\bs{\ta}},z)\circ \Pi^{-1})\btimes (\Pi \circ \mathrm{Y}(e^{-v_k-\bs{q}\cdot\bs{\ta}},z)\circ \Pi^{-1}) \\
 &\quad \left( \Pi(Z^\tor)\btimes\Pi(Z^\tor) \right) \Big)
 \\
=& \sum_{k=1}^\ell\sum_{\bs{q}\in\Z^r} \sum_{\al,\beta\in Q_0 } \res_z \bigg( \ve(v_k+\bs{q}\cdot\bs{\ta},\al)\ve(-v_k-\bs{q}\cdot\bs{\ta},\beta) e^{\xi(\bs{x}^{(k)},z)-\xi(\bs{x'}^{(k)},z)}\cdot \\
&\quad \cdot z^{m+(v_k+\bs{q}\cdot\bs{\ta}|\al)+(-v_k-\bs{q}\cdot\bs{\ta}|\beta)} e^{\sum\limits_{\nu=1}^r q_\nu\left(\xi(\bs{y}^{(\nu)},z)-\xi(\bs{y'}^{(\nu)},z)+w_\nu-w_\nu' \right) } \cdot \\
& \quad \cdot \tau_{\al}\left(\bs{x}-[z^{-1}]^{(k)},\bs{y},\bs{w}\right)   \tau_{\beta}\left(\bs{x}'+[z^{-1}]^{(k)},\bs{y'},\bs{w'}\right)
\bs{u}^{\underline{\al+v_k}}\bs{u'}^{\underline{\beta-v_k} } \bigg)
\\
=& \sum_{k=1}^\ell \sum_{\bs{q}\in\Z^r}  \sum_{\al^\prime,-\beta^\prime\in {Q}_1}\ve(v_k,\al'-v_k)\ve(-v_k,\beta'+v_k)  \cdot \\
& \quad \cdot \res_z \bigg( z^{m+(v_k|\al'-v_k)+(v_k|-\beta'-v_k)} e^{\xi\left(\bs{x}^{(k)}-\bs{x'}^{(k)},z\right)} e^{\sum\limits_{\nu=1}^r q_\nu\left(\xi(\bs{y}^{(\nu)}-\bs{y'}^{(\nu)},z)+w_\nu-w_\nu' \right) }  \cdot \\
& \quad \cdot \tau_{\al'-v_k}\left(\bs{x}-[z^{-1}]^{(k)},\bs{y},\bs{w}\right)  \tau_{\beta'+v_k}\left(\bs{x}'+[z^{-1}]^{(k)},\bs{y'},\bs{w'}\right)
\bigg)
\bs{u}^{\underline{\al'}}\bs{u'}^{\underline{\beta^\prime} },
\end{align*}
where $\bs{q}=(q_1,\dots,q_r)$ and $Q_1$ is given in \eqref{Q1}.
Let us take the coefficients of $\bs{u}^{\underline{\al^\prime} }\bs{u'}^{\underline{\beta^\prime}}$, and do the replacements:
\[
\al^\prime\mapsto \al +v_i, \quad \beta^\prime\mapsto \beta-v_j,
\]
then obtain
\begin{align*}
&  \sum_{k=1}^\ell \sum_{\bs{q}\in\Z^r}  \ve(v_k,\al-\beta+v_i+v_j)  \res_z \bigg(z^{m+(v_k|\al-\beta+v_i+v_j)-2} e^{\xi\left(\bs{x}^{(k)}-\bs{x'}^{(k)},z\right)} \cdot \\
& \quad \cdot e^{\sum\limits_{\nu=1}^r q_\nu\left(\xi(\bs{y}^{(\nu)}-\bs{y'}^{(\nu)},z)+w_\nu-w_\nu' \right) }     \tau_{\al+v_i-v_k}\left(\bs{x}-[z^{-1}]^{(k)},\bs{y},\bs{w}\right) \cdot \\
& \quad \cdot \tau_{\beta-v_j+v_k}\left(\bs{x}'+[z^{-1}]^{(k)},\bs{y'},\bs{w'}\right)
\bigg)=0.
\end{align*}
Doing the following replacements of variables:
\[
\bs{y}\mapsto \bs{y}+\bs{b}, \quad  \bs{w}\mapsto \bs{w}+\bs{c}, \quad \bs{y'}\mapsto \bs{y}-\bs{b}, \quad \bs{w'}\mapsto \bs{w}-\bs{c}
\]
with $\bs{c}=(c_1,\dots,c_\ell)$ being an arbitrary constant vector, we arrive at
\begin{align*}\label{eq-bltor2}
&\sum_{k=1}^{\ell}\sum_{\bs{q}\in\Z^r} \ve(v_k,\al-\beta+v_i+v_j) \res_z \bigg(  z^{(v_k|\al-\beta)+m-2+\de_{i, k}+\de_{j, k}}
e^{\xi(\bs{x}^{(k)}-\bs{x}'^{(k)},z)}  \cdot \\
& \quad \cdot  e^{\sum\limits_{\nu=1}^r 2 q_\nu(\xi(\bs{b}^{(\nu)},z)+c_\nu)}
\tau_{\al+v_i-v_k}\left(\bs{x}-[z^{-1}]^{(k)},\bs{y}+\bs{b},\bs{w}+\bs{c}\right) \cdot \\
& \quad \cdot  \tau_{\beta-v_j+v_k}\left(\bs{x'}+[z^{-1}]^{(k)},\bs{y}-\bs{b},\bs{w}-\bs{c}\right)
\bigg)=0.
\end{align*}
Furthermore, let us apply Lemma~\ref{thm-KIT} with
\begin{align*}
&\Upsilon=\{1,2,\dots,\ell\}, \quad
\ga=k, \quad \zeta=2 c_1, \quad q=q_1, \quad \omega=0,
\end{align*}
and
\begin{align*}
D_{k}(q_1)=& \ve(v_k,\al-\beta+v_i+v_j) \res_z \Bigg(  z^{(v_k|\al-\beta)+m-2+\de_{i, k}+\de_{j, k}}
e^{\xi(\bs{x}^{(k)}-\bs{x}'^{(k)},z)}  \cdot \\
&  \cdot  e^{\sum\limits_{\nu=1}^r 2 q_\nu \xi(\bs{b}^{(\nu)},z)} \exp\left(-\sum_{n>0}\frac{1}{n z^n}\frac{\partial}{\partial x_n^{(k)}}\right) \exp\left(\sum_{n>0}\frac{1}{n z^n}\frac{\pa}{\pa {x'}_n^{(k)}}\right)\Bigg),
\end{align*}
then we obtain
\begin{align*}
0&=\sum_{k=1}^{\ell} D_{k}(0-\frac{1}{2}\frac{\pa}{\pa c_1})\left( \tau_{\al+v_i-v_k}\left(\bs{x},\bs{y}+\bs{b},\bs{w}+\bs{c}\right) \tau_{\beta-v_j+v_k}\left(\bs{x'},\bs{y}-\bs{b},\bs{w}-\bs{c}\right) \right)\Bigg|_{c_1=0}\\
&=\sum_{k=1}^{\ell}\sum_{\bs{q}\in\Z^r} \ve(v_k,\al-\beta+v_i+v_j) \res_z \bigg(  z^{(v_k|\al-\beta)+m-2+\de_{i, k}+\de_{j, k}}
e^{\xi(\bs{x}^{(k)}-\bs{x}'^{(k)},z)}  \cdot \\
& \quad \cdot  e^{\sum\limits_{\nu=2}^r 2 q_\nu(\xi(\bs{b}^{(\nu)},z)+c_\nu)}
\tau_{\al+v_i-v_k}\left(\bs{x}-[z^{-1}]^{(k)},\bs{y}+\bs{b},w_1-\xi(\bs{b}^{(1)},z), w_2+c_2,\cdots w_r+c_r\right) \cdot \\
& \quad \cdot  \tau_{\beta-v_j+v_k}\left(\bs{x'}+[z^{-1}]^{(k)},\bs{y}-\bs{b},w_1+\xi(\bs{b}^{(1)},z),w_2-c_2,\cdots w_r-c_r\right)
\bigg)
.
\end{align*}
For $\nu=2,\dots,r$, we apply Lemma~\ref{thm-KIT} step-by-step with
$
\Upsilon=\{1,2,\dots,\ell\},$
$\ga=k,$ $ \zeta=2 c_\nu,$ $ q=q_\nu,$  $\omega=0,
$ and
\begin{align*}
D_{k}(q_\nu)=& \ve(v_k,\al-\beta+v_i+v_j) \res_z \Bigg(  z^{(v_k|\al-\beta)+m-2+\de_{i, k}+\de_{j, k}}
e^{\xi(\bs{x}^{(k)}-\bs{x}'^{(k)},z)}  \cdot \\
&  \cdot  \exp\left( \sum\limits_{\mu=\nu}^r 2 q_\mu \xi(\bs{b}^{(\mu)},z)
\right) \exp\left(-\sum\limits^{\nu-1}_{\lambda=1}\xi(\bs{b}^{(\lambda)},z) \left.\frac{\pa}{\pa c_\lambda}\right|_{c_\lambda=0} \right) \cdot \\
& \cdot \exp\left(-\sum_{n>0}\frac{1}{n z^n}\frac{\partial}{\partial x_n^{(k)}}\right) \exp\left(\sum_{n>0}\frac{1}{n z^n}\frac{\pa}{\pa {x'}_n^{(k)}}\right)\Bigg)
\end{align*}
acting on the function $\tau_{\al+v_i-v_k}\left(\bs{x},\bs{y}+\bs{b},\bs{w}+\bs{c}\right) \tau_{\beta-v_j+v_k}\left(\bs{x'},\bs{y}-\bs{b},\bs{w}-\bs{c}\right)$. Thus equation \eqref{eq-bltor} is obtained. The theorem is proved.
\end{prfof}

\subsection{Lax equations  associated to $\mathcal{L}^\tor_{r+1}(\fsl_\ell)$}

Similar as in Subsection~\ref{sec-Laxsl}, with the help of tau functions for the vacuum orbit of the Lie group of the toroidal Lie algebra $\fg^\tor\cong\mathcal{L}^\tor_{r+1}(\fsl_\ell)$, we introduce the following matrix pseudo-differential operators
\begin{equation}\label{Paltor}
\bar{P}_\al=\sum_{n\ge0}\bar{P}_{\al,n} \pa^{-n}, \quad \al\in Q_0.
\end{equation}
Here $\pa=\sum_{i=1}^{\ell}\pa/\pa x_{1}^{(i)}$, and the coefficients $\bar{P}_{\al,n}$ are $\ell\times\ell$ matrices with components
\begin{equation}\label{Pnijtor}
\left(\bar{P}_{\al,n}\right)_{i j}=-\ve(v_j,v_i)\res_z \frac{z^{n+\de_{i, j}-2}\tau_{\al+v_i-v_j}\left(\bs{x}-[z^{-1}]^{(j)}, \bs{y}, \bs{w}\right)}{\tau_{\al}(\bs{x}, \bs{y}, \bs{w})}.
\end{equation}
For the same reason as before, the operators $\bar{P}_\al$ are invertible.
\begin{thm}\label{thm-Laxtor}
The bilinear equations \eqref{eq-bltor} are equivalent to
\begin{align}
&\res_z \Bigg( z^m \left.\left(\bar{P}_\al e^{\Xi(\bs{x},z)}\right)\right|_{(\bs{y}, \bs{w})\mapsto (\bs{y}+\bs{b}, \bs{w}-\xi(\bs{b},z))}\cdot\Lambda_{\al-\beta}(z)\cdot \nonumber \\
&\quad \cdot \left.\left( (\bar{P}_\beta^*)^{-1} e^{-\Xi(\bs{x},z)} \right)^T\right|_{(\bs{x},\bs{y}, \bs{w})\mapsto(\bs{x'},\bs{y}-\bs{b}, \bs{w}+\xi(\bs{b},z))}\Bigg)=0, \label{blPtor}
\end{align}
with arbitrary variables $\bs{x}$, $\bs{x}'$, constant vectors $\bs{b}$, $m\in \Z_{\ge0}$, and $\al,\beta\in Q_0$. Moreover, for $1\le i\le \ell$, $1\le\nu\le r$ and $k\geq1$, the operators $\bar{P}_\al$ satisfy
\begin{align}\label{PPinvtor}
 & \left(\bar{P}_\al \pa {\bar{P}_{\al}}^{-1}\right)_{-}=0, \\
 & \frac{\pa \bar{P}_\al}{\pa x^{(i)}_k}=-\left(\bar{P}_\al E_{i i} \pa^k {\bar{P}_{\al}}^{-1}\right)_{-}\bar{P}_\al, \label{Pxtor} \\
 & \frac{\pa \bar{P}_\al}{\pa y^{(\nu)}_k}=\left( \frac{\pa \bar{P}_\al}{\pa w_\nu}  \pa^k {\bar{P}_{\al}}^{-1}\right)_{-}\bar{P}_\al. \label{Pytor}
\end{align}
\end{thm}
\begin{prf}
Equations \eqref{blPtor} are just reformulation of the bilinear equations  \eqref{eq-bltor}, and equations \eqref{PPinvtor} and \eqref{Pxtor} can be derived in the same ways as that for Theorem~\ref{thm-Laxbl} by setting $\bs{b}=0$. Now let us proceed to show \eqref{Pytor}.

Firstly, the bilinear equations \eqref{blPtor} with $\beta=\al$ and $\bs{x'}=\bs{x}$ read
\begin{align*}
&\res_z \Bigg( z^m \left.\left(\bar{P}_\al e^{\Xi(\bs{x},z)}\right)\right|_{(\bs{y}, \bs{w})\mapsto (\bs{y}+\bs{b}, \bs{w}-\xi(\bs{b},z))}\cdot  \\
&\quad \cdot \left.\left( (\bar{P}_\al^*)^{-1} e^{-\Xi(\bs{x},z)} \right)^T\right|_{(\bs{y}, \bs{w})\mapsto(\bs{y}-\bs{b}, \bs{w}+\xi(\bs{b},z))}\Bigg)=0,
\end{align*}
Let $\pa/\pa b^{(\nu)}_k$ act on it and take $\bs{b}=0$, then one has
\begin{align*}
&\res_z \Bigg(  \left(\frac{\pa \bar{P}_\al}{\pa y^{(\nu)}_k}-z^k  \frac{\pa \bar{P}_\al}{\pa w_\nu }\right)\pa^m e^{\Xi(\bs{x},z)} \cdot  \left( (\bar{P}_\al^*)^{-1} e^{-\Xi(\bs{x},z)} \right)^T+ \\
&\quad  +\bar{P}_\al\pa^m  e^{\Xi(\bs{x},z)} \cdot  \left( (\bar{P}_\al^*)^{-1} \left(\frac{\pa \bar{P}_\al^*}{\pa y^{(\nu)}_k}-z^k  \frac{\pa \bar{P}_\al^*}{\pa w_\nu }\right) (\bar{P}_\al^*)^{-1} e^{-\Xi(\bs{x},z)} \right)^T \Bigg)=0,
\end{align*}
namely,
\begin{align}
&\res_\pa \Bigg(  \left(\frac{\pa \bar{P}_\al}{\pa y^{(\nu)}_k}-  \frac{\pa \bar{P}_\al}{\pa w_\nu }\pa^k \right)\pa^m   \bar{P}_\al^{-1}  + \bar{P}_\al\pa^m   \left( \bar{P}_\al^{-1}\frac{\pa \bar{P}_\al}{\pa y^{(\nu)}_k}  \bar{P}_\al^{-1} - \pa^k  \bar{P}_\al^{-1}  \frac{\pa \bar{P}_\al}{\pa w_\nu } \bar{P}_\al^{-1}\right)  \Bigg)=0. \label{eq07}
\end{align}
On the other hand, according to \eqref{PPinvtor} and \eqref{Pxtor}, one has $\pa(\bar{P}_\al)=0$, or equivalently, $[\pa, \bar{P}_\al]=0$. Hence equation \eqref{eq07} leads to
\[
\res_\pa \Bigg(  \left(\frac{\pa \bar{P}_\al}{\pa y^{(\nu)}_k}\bar{P}_\al^{-1}-  \frac{\pa \bar{P}_\al}{\pa w_\nu }\pa^k \bar{P}_\al^{-1} \right)\pa^m  + \pa^m   \left(\frac{\pa \bar{P}_\al}{\pa y^{(\nu)}_k}  \bar{P}_\al^{-1} - \frac{\pa \bar{P}_\al}{\pa w_\nu } \pa^k \bar{P}_\al^{-1}\right)  \Bigg)=0.
\]
It holds for all $m\ge0$, thus we arrive at
\[
\left(\frac{\pa \bar{P}_\al}{\pa y^{(\nu)}_k}\bar{P}_\al^{-1}-  \frac{\pa \bar{P}_\al}{\pa w_\nu }\pa^k \bar{P}_\al^{-1} \right)_-=0,
\]
which implies \eqref{Pytor}. The theorem is proved.
\end{prf}

From \eqref{PPinvtor} and \eqref{Pxtor} it follows that
\[
\sum_{i=1}^\ell \frac{\pa \bar{P}_\al}{\pa x_{k}^{(i)} }=0, \quad \al\in Q_0, \ k\ge1.
\]
For the same reason as before, it can be assumed that solutions of the bilinear equation \eqref{eq-bltor} satisfy
\begin{equation}\label{patau}
\sum_{i=1}^\ell \frac{\pa \tau_\al}{\pa x_{k}^{(i)}}=0, \quad \al\in Q_0, \ k\ge1.
\end{equation}

By comparing equations \eqref{PPinvtor}--\eqref{Pytor} with \eqref{PmPinv}--\eqref{Pxik}, one sees that the integrable hierarchy \eqref{eq-bltor} associated to $\mathcal{L}^\tor_{r+1}(\fsl_\ell)$ is an extension of the $(1,1,\dots,1)$-reduction of the $\ell$-component KP hierarchy.

\begin{exa}
Let us consider a particular case of the bilinear equation \eqref{eq-bltor}. We take $\ell=2$, $i=j=2$, $\al=p(v_1-v_2)$ and $\beta=q(v_1-v_2)$ with $p,q\in\Z$, and write $\tau_{p}=\tau_{p(v_1-v_2)}$ for short, then equation \eqref{eq-bltor} reads: for $m\ge0$,
\begin{align*}
&\ve(v_1,(p-q)(v_1-v_2))\res_z \Bigg( z^{(v_1|(p-q)(v_1-v_2))+m-2} e^{\xi(\bs{x}^{(1)}-\bs{x'}^{(1)},z)} \cdot \\
& \quad \cdot
\tau_{p-1}\left(\bs{x}-[z^{-1}]^{(1)},\bs{y}+\bs{b},\bs{w}-\xi(\bs{b},z)\right)
 \tau_{q+1}\left(\bs{x'}+[z^{-1}]^{(1)},\bs{y}-\bs{b},\bs{w}+\xi(\bs{b},z)\right) \Bigg)
+\\
&\quad +\ve(v_2,(p-q)(v_1-v_2))\res_z \Bigg( z^{(v_2|(p-q)(v_1-v_2))+m} e^{\xi(\bs{x}^{(2)}-\bs{x'}^{(2)},z)} \cdot \\
& \quad \cdot
\tau_{p}\left(\bs{x}-[z^{-1}]^{(2)},\bs{y}+\bs{b},\bs{w}-\xi(\bs{b},z)\right)
 \tau_{q}\left(\bs{x'}+[z^{-1}]^{(1)},\bs{y}-\bs{b},\bs{w}+\xi(\bs{b},z)\right) \Bigg) =0.
\end{align*}
Namely,
\begin{align*}
& \res_z \Bigg( z^{m+p-q-2} e^{\xi(\bs{x}^{(1)}-\bs{x'}^{(1)},z)}
\tau_{p-1}\left(\bs{x}-[z^{-1}]^{(1)},\bs{y}+\bs{b},\bs{w}-\xi(\bs{b},z)\right)
\cdot \\
& \quad \cdot\tau_{q+1}\left(\bs{x'}+[z^{-1}]^{(1)},\bs{y}-\bs{b},\bs{w}+\xi(\bs{b},z)\right) \Bigg)
+(-1)^{p-q}\res_z \Bigg( z^{m+p-q} e^{\xi(\bs{x}^{(2)}-\bs{x'}^{(2)},z)} \cdot \\
& \quad \cdot
\tau_{p}\left(\bs{x}-[z^{-1}]^{(2)},\bs{y}+\bs{b},\bs{w}-\xi(\bs{b},z)\right)
 \tau_{q}\left(\bs{x'}+[z^{-1}]^{(1)},\bs{y}-\bs{b},\bs{w}+\xi(\bs{b},z)\right) \Bigg) =0.
\end{align*}
One introduces variables $\bs{t}=(t_1,t_2,\dots)$ and  $\bs{\tilde t}=({\tilde t}_1,{\tilde t}_2,\dots)$ by
\[
t_k=x_k^{(1)}-x_k^{(2)}, \quad {\tilde t}_k=x_k^{(1)}+x_k^{(2)}.
\]
Then the tau functions can be represented as follows
\begin{equation}\label{taut0}
\tau_{p}\left(\bs{x}^{(1)}, \bs{x}^{(2)} ,\bs{y},\bs{w}\right)=\tau_{p}\left(\frac{1}{2}(\bs{t}+\bs{\tilde t}), \frac{1}{2}(\bs{\tilde t}-\bs{t}), \bs{y},\bs{w}\right).
\end{equation}
The equalities \eqref{patau} means that the right hand side does not depend on $\bs{\tilde t}$. Let ${\tilde\tau}_{p}(\bs{t},\bs{y},\bs{w})$ denote the right hand side of \eqref{taut0}, then one has, for $m\ge0$,
\begin{align*}
& \res_z \Bigg( z^{m+p-q-2} e^{\xi(\bs{t}-\bs{t'},z)/2}
{\tilde \tau}_{p-1}\left(\bs{t}-[z^{-1}],\bs{y}+\bs{b},\bs{w}-\xi(\bs{b},z)\right)
\cdot \\
& \quad \cdot {\tilde \tau}_{q+1}\left(\bs{t'}+[z^{-1}],\bs{y}-\bs{b},\bs{w}+\xi(\bs{b},z)\right) \Bigg)
+(-1)^{p-q}\res_z \Bigg( z^{m+p-q} e^{\xi(\bs{t'}-\bs{t},z)/2} \cdot \\
& \quad \cdot
{\tilde \tau}_{p}\left(\bs{t}+[z^{-1}],\bs{y}+\bs{b},\bs{w}-\xi(\bs{b},z)\right)
 {\tilde \tau}_{q}\left(\bs{t'}-[z^{-1}],\bs{y}-\bs{b},\bs{w}+\xi(\bs{b},z)\right) \Bigg) =0.
\end{align*}
This equation coincides with equation~(4.41) in \cite{KIT} via the identification
\[
{\tilde \tau}_{p}\left(\bs{t},\bs{y}+\bs{b},\bs{w}-\xi(\bs{b},z)\right) \
\leftrightarrow \ \tau_{0,0}^{p,-p}(\underline{x}, \underline{y}-\underline{b}_z).
\]
Note that equation~(4.41) in \cite{KIT} is derived starting from a Fock space of free fermions, whose solutions with $r=1$ are also related to tau functions of the (2+1)-dimensional Toda lattice hierarchy \cite{Ogawa2008}. What is more, some equations of the components of $\bar{P}_{v_1-v_2,\,1}$ (recall \eqref{Pnijtor}) have been written down explicitly in \cite{KIT}, which can be reduced to the a $(2 + 1)$-dimensional extension of the nonlinear Schr\"{o}dinger equation.
\end{exa}

\begin{exa}
Let us take $\ell=3$, $i=j=1$, $\al=p_1v_1+p_2v_2-(p_1+p_2)v_3$ and $\beta=q_1v_1+q_2v_2-(q_1+q_2)v_3$ with $p_i,q_i\in\Z$, and write $\tau_{p_1,p_2}=\tau_{p_1v_1+p_2v_2-(p_1+p_2)v_3}$ for short,  then the bilinear equation \eqref{eq-bltor} reads: for $m\ge0$,
 \begin{align*}
&{\rm Res}_z \Big( z ^{p_1-q_1+m}e^{\xi\left(\bs{x}^{(1)}-\bs{x'}^{(1)},z\right)} \tau_{p_1,p_2}(\bs{x}-[z^{-1}]^{(1)},\bs{y}+\bs{b},\bs{w}-\xi(\bs{b},z))\cdot \\
&\quad \cdot\tau_{q_1,q_2}(\bs{x'}+[z^{-1}]^{(1)},\bs{y}-\bs{b},\bs{w}+\xi(\bs{b},z)) +  \\
&\quad +(-1)^{ p_2-q_2} z ^{p_2-q_2+m-2} e^{\xi\left(\bs{x}^{(2)}-\bs{x'}^{(2)},z\right)} \tau_{p_1+1,p_2-1}(\bs{x}-[z^{-1}]^{(2)},\bs{y}+\bs{b},\bs{w}-\xi(\bs{b},z))\cdot  \\ &\quad\cdot\tau_{q_1-1,q_2+1}(\bs{x'}+[z^{-1}]^{(2)},\bs{y}+\bs{b},\bs{w}-\xi(\bs{b},z)) + \\
&\quad + (-1)^{p_1-q_1}  z ^{q_1-p_1+q_2-p_2+m-2}  e^{\xi\left(\bs{x}^{(3)}-\bs{x'}^{(3)},z\right)} \tau_{p_1+1,p_2}(\bs{x}-[z^{-1}]^{(3)},\bs{y}+\bs{b},\bs{w}-\xi(\bs{b},z))\cdot\nonumber \\ &\quad \cdot\tau_{q_1-1,q_2}(\bs{x'}+[z^{-1}]^{(3)},\bs{y}-\bs{b},\bs{w}+\xi(\bs{b},z))\Big)=0.
\end{align*}
For example, following are some explicit Hirota bilinear equations:
\begin{align*}
&D_{x^{(1)}_1}D_{x^{(2)}_1}\tau_{p_1,p_2}\cdot\tau_{p_1,p_2}-2\tau_{p_1+1,p_2-1}\cdot\tau_{p_1-1,p_2+1} =0,\\
&D_{x^{(1)}_1}D_{x^{(1)}_2}\tau_{p_1,p_2}\cdot\tau_{p_1,p_2}-2D_{x^{(1)}_1}\left(
        \tau_{p_1+1,p_2-1}\tau_{p_1-1,p_2+1}+\tau_{p_1+1,p_2}\cdot\tau_{p_1-1,p_2}\right)=0,\\
&\left(D_{y^{(1)}_1}D_{x^{(1)}_1}-\frac{1}{2}D_{w_1}D_{x^{(2)}_1}\right) \tau_{p_1,p_2}\cdot\tau_{p_1,p_2}\\
        &\quad
        -D_{w_1}\left(\tau_{p_1+1,p_2-1}\cdot\tau_{p_1-1,p_2+1} +\tau_{p_1+1,p_2}\cdot\tau_{p_1-1,p_2}\right)=0,\\
&\left(\frac{1}{2}D_{y^{(1)}_1}D_{x^{(2)}_1}-\frac{1}{6}D_{w_1}D^3_{x^{(1)}_1} -\frac{1}{3}D_{w_1}D_{x^{(1)}_3}\right)\tau_{p_1,p_2}\cdot\tau_{p_1,p_2}\\
         &\quad +\left(D_{y^{(1)}_1}-D_{w_1}D_{x^{(2)}_1}\right)
        \tau_{p_1+1,p_2-1}\cdot\tau_{p_1-1,p_2+1} \\
        &\quad  +\left(D_{y^{(1)}_1}-D_{w_1}D_{x^{(3)}_1}\right)\tau_{p_1+1,p_2}\cdot\tau_{p_1-1,p_2}=0,\\
&\left(\frac{1}{2}D_{w_1}D^2_{x^{(1)}_1}D_{x^{(2)}_1}\right)\tau_{p_1,p_2}\cdot\tau_{p_1,p_2}+D_{w_1}D_{x^{(2)}_1}\tau_{p_1+1,p_2}\cdot\tau_{p_1-1,p_2}\\
        &\quad +\left(2D_{y^{(1)}_1}-D_{w_1}D_{x^{(2)}_1}\right)
        \tau_{p_1+1,p_2-1}\cdot\tau_{p_1-1,p_2+1}=0.
\end{align*}
Here $D_\zeta$ stands for the Hirota derivation, namely,
\[
D_\zeta f\cdot g=\left.\frac{\pa }{\pa u}\right|_{u=0}f(\zeta+u)g(\zeta-u).
\]

In particular, let us assume $\tau_\al$ to be complex functions and write (recall \eqref{Pnijtor})
\[
\bar{P}_{v_1-v_2,1}=\left(
                      \begin{array}{ccc}
                        f & u & w \\
                       \bar{u} & g & v \\
                        \bar{w} & \bar{v} & -f-g \\
                      \end{array}
                    \right);
\]
we also assume the variables $x,\,t,\,y,\,s$ to be real ones such that ($\mathbf{i}=\sqrt{-1}$)
\[
\pa_x=\pa_{x_1^{(1)}}-\pa_{x_1^{(2)}}, \quad \mathbf{i} \pa_t =\pa_{x_2^{(1)}}-\pa_{x_2^{(2)}}, \quad \pa_y=\pa_{w_1}, \quad \mathbf{i}\pa_s=\pa_{y_1^{(1)} }.
\]
Then, starting from \eqref{PPinvtor} and \eqref{Pxtor} a straightforward calculation leads to the following equations:
\begin{align}
2\mathbf{i}u_t&=u_{x x}- \bar{v} w_x + w \bar{v}_x+u(8|u|^{2}+2|v|^2+2|w|^{2}), \label{eqsl3-1}\\
-\mathbf{i}v_t&=v_{x x}-2\bar{u}w_x-w \bar{u}_x+v(2|u|^2+2|v|^{2}-|w|^{2}),\\
\mathbf{i} w_t&=w_{x x}+2 u v_x+v u_x +w(2|u|^2-|v|^2 + 2|w|^{2}).  \label{eqsl3-3}
\end{align}
Moreover, taking \eqref{Pytor} into account one derives
\begin{align}
2\mathbf{i}u_s&=u_{x y}- \bar{v} w_y + w \bar{v}_y+u\int(8|u|^{2}+2|v|^2+2|w|^{2})_y\mathrm{d}x,  \label{eqsl3-4}\\
-\mathbf{i}v_s&=v_{x y}-2\bar{u}w_y-w \bar{u}_y+v\int(2|u|^2+2|v|^{2}-|w|^{2})_y\mathrm{d}x,\\
\mathbf{i} w_s&=w_{x y}+2 u v_y+v u_y +w\int(2|u|^2-|v|^2 + 2|w|^{2})_y\mathrm{d}x.  \label{eqsl3-6}
\end{align}
One observes that these three equations can be considered as some generalization of the (2+1)-dimensional nonlinear Schr\"odinger equation derived from $\mathcal{L}_{2}^{\tor}(\fsl_2)$ (see \cite{KIT}), and they are recast to equations \eqref{eqsl3-1}--\eqref{eqsl3-3} from reduction of the $3$-component KP hierarchy whenever $y=x$.
\end{exa}

\section{Concluding remarks}

In this paper we apply an explicit homogeneous realization of the toroidal Lie algebra $\mathcal{L}^\tor_{r+1}(\fsl_\ell)$ via lattice vertex algebras to construct an integrable hierarchy of Hirota bilinear equations. This hierarchy is recast to Lax equations of pseudo-differential operators with matrix coefficients, and shown to be an extension of the $(1,1,\dots,1)$-reduced $\ell$-component KP hierarchy. Our results agree with that on the hierarchy of (2+1)-dimensional nonlinear Schr\"odinger equation associated to $\mathcal{L}^\tor_{2}(\fsl_2)$ obtained in \cite{KIT} with a different method.

In our construction, we write down a basis of a subalgebra of some lattice vertex algebra that is isomorphic with the toroidal Lie algebra $\mathcal{L}^\tor_{r+1}(\fsl_\ell)$. It is known that every simple Lie algebra can be regarded as a subalgebra of $\fsl_\ell$, hence in principle we have a homogeneous realization of the toroidal Lie algebra for an arbitrary simple Lie algebra. Based on this fact, it is natural to study the corresponding integrable hierarchies and their relationship with integrable hierarchies of KP type. What is more, in the framework of lattice vertex algebras, the twisted representations of toroidal Lie algebras were studied thoroughly by Bakalov and Kirk \cite{Bakalov2021}. To our best knowledge, it is still open how to generalize our construction to twisted cases. These questions will be considered elsewhere.

{\bf Acknowledgments.}
{\noindent \small The authors thank Fulin Chen and Kanehisa Takasaki for helpful discussions; they also thank the anonymous referees for their encouragement and very useful suggestions. This work is partially supported by National Key R\&D Program of China 2023YFA10098001, NSFC No.\,12471243 and Guangzhou S\&T Program No. SL2023A04J01542. }

\begin{appendices}
\section*{Appendix}
\section{Definition of toroidal Lie algebras}
Let us recall the definition of toroidal Lie algebras. Let $\fg_0$ be a semisimple complex Lie algebra with the standard nondegenerate symmetric bilinear form $(\cdot|\cdot)$. Given a nonnegative integer $r$, consider the ring $\mathcal{R}$ of Laurent polynomials in $r+
1$ parameters $t_0, t_1, \dots, t_r$, that is,
\[
\mathcal{R}=\C\left[t_0^{\pm 1},\ t_1^{\pm 1},\dots, t_r^{\pm 1}\right].
\]
Let $K_0, K_1,\dots, K_r$ be linearly independent vectors of a certain linear space, and
\[
\mathcal{K}=\left.\left(\bigoplus_{\nu=0}^r \mathcal{R}K_\nu \right)\right/\mathrm{span}_\C\left\{
\sum_{\nu=0}^r m_\nu t_0^{m_0} t_1^{m_1}\dots t_r^{m_r}K_\nu\mid m_0,m_1,\dots,m_r\in\Z \right\}.
\]
The (untwisted) toroidal Lie algebra associated to $\fg_0$ with $r+1$ parameters is
\[
\mathcal{L}_{r+1}^{\mathrm{tor}}(\fg_0)=(\fg_0\otimes\mathcal{R})\oplus\mathcal{K},
\]
whose bracket is defined by
\begin{align*}
&[\mathcal{L}_{r+1}^\tor(\fg_0),\mathcal{K}]=0, \\
&[X\otimes t_0^{m_0}\dots t_r^{m_r}, Y\otimes t_0^{n_0}\dots t_r^{n_r}] \\
=& [X,Y]\otimes t_0^{m_0+n_0}\dots t_r^{m_r+n_r}+(X|Y)\sum_{\nu=0}^r \overline{m_\nu t_0^{m_0+n_0}\dots t_r^{m_r+n_r}K_\nu}\, ,
\end{align*}
with $X,Y\in\fg_0$; $m_\nu, n_\nu\in\Z$; the notation $\overline{C}$ standing for the equivalence class in the quotient space $\mathcal{K}$.

The notion  of toroidal Lie algebra is a natural generalization of affine Lie algebra. More exactly, when $r=0$, denote $t_0=s$ and $K=\overline{K_0}\in\mathcal{K}$. In the toroidal Lie algebra $\mathcal{L}_1^\tor(\fg_0)$, the center is
\[
\mathcal{K}=\left.\C[s^{\pm 1}]K_0\right./\mathrm{span}_\C\left\{m s^m K_0\mid m\in\Z \right\}=\C K,
\]
and the bracket is (cf. \eqref{aff})
\[
[X\otimes s^m, Y\otimes s^n]=[X, Y]\otimes s^{m+n}+(X|Y)\de_{m,-n}m K.
\]
It means that $\mathcal{L}_1^\tor(\fg_0)$ is just the untwisted affine Lie algebra associated to $\fg_0$ \cite{Kac1990}.

\section{Proof of Proposition~\ref{thm-hep}}

We proceed to verify Proposition~\ref{thm-hep}, with the help of the equalities \eqref{ambnbra},  \eqref{Gabl} and \eqref{Gaep}. It is straightforward to verify
\begin{align*}
&\left[ h_{i,m}^{\bs{p}},h_{j,n}^{\bs{q}}\right] \\
=& \sum_{k\ge0}\binom{m}{k}\left( \left( (v_i-v_{i+1})s^{-1}\otimes e^{\bs{p}\cdot\bs{\ta}}\right)_{(k)}  \left( (v_j-v_{j+1})s^{-1}\otimes e^{\bs{q}\cdot\bs{\ta}}\right) \right)_{(m+n-k)}
\\
=&  \sum_{k\ge0}\binom{m}{k}\res_z\Bigg( z^k :\sum_{t\in\Z} (v_i-v_{i+1})_{(-t)}z^{t-1} \cdot \\ &\quad\cdot e^{\bs{p}\cdot\bs{\ta}} z^{ (\bs{p}\cdot\bs{\ta})_{(0)}}
\exp\left(\sum_{d>0}(\bs{p}\cdot\bs{\ta})_{(-d)}\frac{z^d}{d}\right) \exp\left(\sum_{d<0}(\bs{p}\cdot\bs{\ta})_{(-d)}\frac{z^d}{d}\right):  \\
&\quad \left( (v_j-v_{j+1})s^{-1}\otimes e^{\bs{q}\cdot\bs{\ta}}\right) \Bigg)_{(m+n-k)}
\\
=&\sum_{k\ge0}\binom{m}{k}\res_z\Bigg( z^k \sum_{t\ge1} (v_i-v_{i+1})_{(-t)}z^{t-1}\cdot
\\
 &\quad \cdot \exp\left(\sum_{d>0}(\bs{p}\cdot\bs{\ta})_{(-d)}\frac{z^d}{d}\right) \left( (v_j-v_{j+1})s^{-1}\otimes e^{(\bs{p}+\bs{q})\cdot\bs{\ta}} \right) \Bigg)_{(m+n-k)}+
\\
&\quad + \sum_{k\ge0}\binom{m}{k}\res_z\Bigg( z^k  e^{\bs{p}\cdot\bs{\ta}} z^{ (\bs{p}\cdot\bs{\ta})_{(0)}} \exp\left(\sum_{d>0}(\bs{p}\cdot\bs{\ta})_{(-d)}\frac{z^d}{d}\right) \exp\left(\sum_{d<0}(\bs{p}\cdot\bs{\ta})_{(-d)}\frac{z^d}{d}\right) \cdot
\\
 &\quad \cdot \sum_{t\le0} (v_i-v_{i+1})_{(-t)}z^{t-1} \left( (v_j-v_{j+1})s^{-1}\otimes e^{\bs{q}\cdot\bs{\ta}} \right)
 \Bigg)_{(m+n-k)}
 \\
=&0 + \sum_{k\ge0}\binom{m}{k}\res_z\Big( z^{k-2} \left(1+ (\bs{p}\cdot\bs{\ta})_{(-1)}z\right)\cdot
\\
 &\quad \cdot (v_i-v_{i+1})_{(1)} \left( (v_j-v_{j+1})s^{-1}\otimes e^{(\bs{p}+\bs{q})\cdot\bs{\ta}} \right)
 \Big)_{(m+n-k)}
 \\
=&    \binom{m}{1} (v_i-v_{i+1})_{(1)} \left( (v_j-v_{j+1})s^{-1}\otimes e^{(\bs{p}+\bs{q})\cdot\bs{\ta}} \right)_{(m+n-1)}  \\
 &\quad +\binom{m}{0} (\bs{p}\cdot\bs{\ta})_{(-1)}(v_i-v_{i+1})_{(1)}\left( (v_j-v_{j+1})s^{-1}\otimes e^{(\bs{p}+\bs{q})\cdot\bs{\ta}} \right)_{(m+n)} \\
=& (2\de_{i,j}-\de_{i,j+1}-\de_{i+1,j})\left( \left( m \otimes e^{(\bs{p}+\bs{q})\cdot\bs{\ta}} \right)_{(m+n-1)}  +  \left(\bs{p}\cdot\bs{\ta} s^{-1}\otimes e^{(\bs{p}+\bs{q})\cdot\bs{\ta}} \right)_{(m+n)} \right)\\
 =& (2\de_{i,j}-\de_{i,j+1}-\de_{i+1,j})\left( m \ka_{0,m+n}^{\bs{p}+\bs{q}}+\sum_{\nu=1}^r p_\nu \ka_{\nu,m+n}^{\bs{p}+\bs{q}} \right),
\end{align*}
\begin{align*}
&\left[ h_{i,m}^{\bs{p}}, e_{k l,n}^{\bs{q}}\right] \\
=&\ep_{l k} \sum_{j\ge0}\binom{m}{j}\left( \left( (v_i-v_{i+1})s^{-1}\otimes e^{\bs{p}\cdot\bs{\ta}}\right)_{(j)}  \left(  1\otimes e^{v_k-v_l+\bs{q}\cdot\bs{\ta}}\right) \right)_{(m+n-j)}
\\
=& \ep_{l k} \sum_{j\ge0}\binom{m}{j}\res_z\Bigg( z^j :\sum_{t\in\Z} (v_i-v_{i+1})_{(-t)}z^{t-1} \cdot \\
&\quad \cdot e^{\bs{p}\cdot\bs{\ta}} z^{ (\bs{p}\cdot\bs{\ta})_{(0)}}
\exp\left(\sum_{d>0}(\bs{p}\cdot\bs{\ta})_{(-d)}\frac{z^d}{d}\right) \exp\left(\sum_{d<0}(\bs{p}\cdot\bs{\ta})_{(-d)}\frac{z^d}{d}\right):  \\
&\quad \left(  1\otimes e^{v_k-v_l+\bs{q}\cdot\bs{\ta}} \right) \Bigg)_{(m+n-j)}
\\
=& \ep_{l k} \sum_{j\ge0}\binom{m}{j}\res_z\Bigg( z^j \sum_{t\ge1} (v_i-v_{i+1})_{(-t)}z^{t-1}
\\
 &\quad \exp\left(\sum_{d>0}(\bs{p}\cdot\bs{\ta})_{(-d)}\frac{z^d}{d}\right) \left(  1\otimes e^{v_k-v_l+(\bs{p}+\bs{q})\cdot\bs{\ta}}  \right) \Bigg)_{(m+n-j)}+
\\
&\quad + \ep_{l k} \sum_{j\ge0}\binom{m}{j}\res_z\Bigg( z^j  e^{\bs{p}\cdot\bs{\ta}} z^{ (\bs{p}\cdot\bs{\ta})_{(0)}}
\exp\left(\sum_{d>0}(\bs{p}\cdot\bs{\ta})_{(-d)}\frac{z^d}{d}\right) \cdot \\
&\quad\cdot
\exp\left(\sum_{d<0}(\bs{p}\cdot\bs{\ta})_{(-d)}\frac{z^d}{d}\right)   \sum_{t\le0} (v_i-v_{i+1})_{(-t)}z^{t-1} \left(  1\otimes e^{v_k-v_l+\bs{q}\cdot\bs{\ta}} \right) \Bigg)_{(m+n-j)}
\\
=& 0+ \ep_{l k}  \binom{m}{0}  (v_i-v_{i+1})_{(0)} \left(  1\otimes e^{v_k-v_l+(\bs{p}+\bs{q})\cdot\bs{\ta}}  \right) \Bigg)_{(m+n)}  \\
=& \ep_{l k}  (v_i-v_{i+1}|v_k-v_l) \left(  1\otimes e^{v_k-v_l+(\bs{p}+\bs{q})\cdot\bs{\ta}}  \right) \Bigg)_{(m+n)}   \\
 =&(\de_{i, k}+\de_{i+1,l}-\de_{i, l}-\de_{i+1,k}) e_{k l,m+n}^{\bs{p}+\bs{q}}.
\end{align*}

Moreover, we have
\begin{align*}
&\left[ e_{k l,m}^{\bs{p}},e_{k' l',n}^{\bs{q}}\right]
\\
=&\ep_{l k}\ep_{l' k'}\sum_{j\ge0}\binom{m}{j}\left( \left(1\otimes e^{v_k-v_l+\bs{p}\cdot\bs{\ta}}\right)_{(j)}  \left(  1\otimes e^{v_k'-v_l'+\bs{q}\cdot\bs{\ta}}\right) \right)_{(m+n-j)}
\\
=&\ep_{l k}\ep_{l' k'}\sum_{j\ge0}\binom{m}{j} \res_z\Bigg( z^j  e^{v_k-v_l+\bs{p}\cdot\bs{\ta}} z^{(v_k-v_l+\bs{p}\cdot\bs{\ta})_{(0)}} \exp\left(\sum_{d>0} (v_k-v_l+\bs{p}\cdot\bs{\ta})_{(-d)}\frac{z^d}{d} \right)\cdot \\
 &\quad \cdot
 \exp\left(\sum_{d<0} (v_k-v_l+\bs{p}\cdot\bs{\ta})_{(-d)}\frac{z^d}{d} \right)\left(  1\otimes e^{v_{k'}-v_{l'}+\bs{q}\cdot\bs{\ta}}\right) \Bigg)_{(m+n-j)}
 \\
=&\ep_{l k}\ep_{l' k'}\ve(v_k-v_l+\bs{p}\cdot\bs{\ta},v_{k'}-v_{l'}+\bs{q}\cdot\bs{\ta}) \sum_{j\ge0}\binom{m}{j}\res_z \Bigg( z^{j+(v_k-v_l+\bs{p}\cdot\bs{\ta}|v_{k'}-v_{l'}+\bs{q}\cdot\bs{\ta})}\cdot  \\
 &\quad \cdot
 \exp\left(\sum_{d>0} (v_k-v_l+\bs{p}\cdot\bs{\ta})_{(-d)}\frac{z^d}{d} \right)  \left(  1\otimes e^{v_k-v_l+v_{k'}-v_{l'}+(\bs{p}+\bs{q})\cdot\bs{\ta}}\right) \Bigg)_{(m+n-j)}
 \\
=&\ep_{l k}\ep_{l' k'}\ve(v_k-v_l,v_{k'}-v_{l'}) \binom{m}{0} \res_z \Bigg(z^{(v_k-v_l|v_{k'}-v_{l'})} \cdot  \\
 &\quad \cdot \left(1+(v_k-v_l+\bs{p}\cdot\bs{\ta})_{(-1)}z \right)  \left(  1\otimes e^{v_k-v_l+v_{k'}-v_{l'}+(\bs{p}+\bs{q})\cdot\bs{\ta}}\right) \Bigg)_{(m+n)}+ \\
 &\quad + \de_{k, l'}\de_{l, k'}\ep_{l k}\ep_{k l}\ve(v_k-v_l,v_l-v_k) \binom{m}{1} \left(  1\otimes e^{(\bs{p}+\bs{q})\cdot\bs{\ta}}\right) \Bigg)_{(m+n-1)}
\\
=&\left\{
\begin{array}{cl}
 \left((v_k-v_l)s^{-1}\otimes e^{(\bs{p}+\bs{q})\cdot\bs{\ta} }\right)_{(m+n)}  + m \ka_{0,m+n}^{\bs{p}+\bs{q}}
  +\sum_{\nu=1}^r p_\nu \ka_{\nu,m+n}^{\bs{p}+\bs{q}} ,  &  k=l',\ l=k'; \\
  e_{k l', m+n}^{\bs{p}+\bs{q}},  &  k\ne l',\ l=k'; \\
 -e_{k' l, m+n}^{\bs{p}+\bs{q}},  & k=l', \ l\ne k'; \\
 0,  &  \hbox{ else},
\end{array}
\right.
\end{align*}
which confirms \eqref{epep}. Note that in the last equality it is used the commutation relations \eqref{eekl}.

Finally, suppose
\[
\beta\in Q, \quad g\in\C\left[ v_i s^{-1}, \ta_\nu s^{-1}\mid 1\le i\le \ell, \, 1\le\nu\le r\right].
\]
Then for any $1\le\nu\le r$, one has
\begin{align*}
&\left[ \ka_{\nu,m}^{\bs{p}}, \left(g\otimes e^{\beta+\bs{q}\cdot\bs{\ta}}\right)_{(n)}\right]
\\
=&\sum_{j\ge0}\binom{m}{j}\left( \left(\ta_\nu s^{-1}\otimes e^{\bs{p}\cdot\bs{\ta}}\right)_{(j)}  \left(  g\otimes e^{\beta+\bs{q}\cdot\bs{\ta}} \right) \right)_{(m+n-j)}
\\
=&\sum_{j\ge0}\binom{m}{j} \res_z \Bigg(z^j :\sum_{t\in\Z} (\ta_\nu)_{(-t)} z^{t-1} \cdot \\
&\quad \cdot   e^{\bs{p}\cdot\bs{\ta}} z^{(\bs{p}\cdot\bs{\ta})_{(0)}} \exp\left(\sum_{d>0} (\bs{p}\cdot\bs{\ta})_{(-d)}\frac{z^d}{d} \right)
 \exp\left(\sum_{d<0} (\bs{p}\cdot\bs{\ta})_{(-d)}\frac{z^d}{d} \right): \left(  g\otimes e^{\beta+\bs{q}\cdot\bs{\ta}} \right) \Bigg)_{(m+n-j)}
 \\
=&\ve(\bs{p}\cdot\bs{\ta},\beta+\bs{q}\cdot\bs{\ta}) \sum_{j\ge0}\binom{m}{j}\res_z\Bigg(  z^{j+(\bs{p}\cdot\bs{\ta}|\beta+\bs{q}\cdot\bs{\ta})} \sum_{t\ge1} (\ta_\nu)_{(-t)} z^{t-1}   \\
 &\quad \exp\left(\sum_{d>0} (\bs{p}\cdot\bs{\ta})_{(-d)}\frac{z^d}{d} \right) \left(  g\otimes e^{\beta+(\bs{p}+\bs{q})\cdot\bs{\ta}} \right) \Bigg)_{(m+n-j)}+
 \\
 &\quad + \sum_{j\ge0}\binom{m}{j} \res_z \Bigg(z^j e^{\bs{p}\cdot\bs{\ta}} z^{(\bs{p}\cdot\bs{\ta})_{(0)}} \exp\left(\sum_{d>0} (\bs{p}\cdot\bs{\ta})_{(-d)}\frac{z^d}{d} \right)
 \exp\left(\sum_{d<0} (\bs{p}\cdot\bs{\ta})_{(-d)}\frac{z^d}{d} \right)\cdot \\
&\quad \cdot   \sum_{t\le0} (\ta_\nu)_{(-t)} z^{t-1}  \left(  g\otimes e^{\beta+\bs{q}\cdot\bs{\ta}} \right) \Bigg)_{(m+n-j)}
 \\
=&0 + \sum_{j\ge0}\binom{m}{j} \res_z \Bigg(z^{j-1} e^{\bs{p}\cdot\bs{\ta}} z^{(\bs{p}\cdot\bs{\ta})_{(0)}} \exp\left(\sum_{d>0} (\bs{p}\cdot\bs{\ta})_{(-d)}\frac{z^d}{d} \right)\cdot \\
&\quad \cdot  (\ta_\nu)_{(0)}  \left(  g\otimes e^{\beta+\bs{q}\cdot\bs{\ta}} \right) \Bigg)_{(m+n-j)}
 \\
=& \binom{m}{0} (\ta_\nu| \beta+ \bs{q}\cdot\bs{\ta} ) \left(  g\otimes e^{\beta+(\bs{p}+\bs{q})\cdot\bs{\ta}} \right)_{(m+n)} =0.
\end{align*}
In the same way, it can be verified
\[
\left[ \ka_{0,m}^{\bs{p}}, \left(g\otimes e^{\beta+\bs{q}\cdot\bs{\ta}}\right)_{(n)}\right]=0.
\]
Therefore, Proposition~\ref{thm-hep} is proved.

\end{appendices}

\end{document}